\documentstyle[preprint,aps]{revtex}
\tighten
\draft
\title{Condensate Fluctuations in Trapped Bose Gases: \\
    Canonical vs.\ Microcanonical Ensemble~\footnote{\em
    Dedicated to Professor Dr.\ Siegfried Gro{\ss}mann on the occasion
    of his 68-th birthday}
    }
\author{Martin Holthaus and Eva Kalinowski}
\address{Fachbereich Physik der Philipps-Universit\"at,
    Renthof 6, D-35032 Marburg, Germany}
\author{Klaus Kirsten}
\address{Universit\"at Leipzig, Institut f\"ur Theoretische Physik,
    Augustusplatz 10, D-04109 Leipzig, Germany}
\date{April 15, 1998}
\narrowtext
\newcommand{\nats}{\mbox{${\rm I\!N }$}}

\begin{document}

\maketitle

\begin{abstract}
We study the fluctuation of the number of particles in ideal Bose--Einstein
condensates, both within the canonical and the microcanonical ensemble.
Employing the Mellin--Barnes transformation, we derive simple expressions
that link the canonical number of condensate particles, its fluctuation,
and the difference between canonical and microcanonical fluctuations to the
poles of a Zeta function that is determined by the excited single-particle
levels of the trapping potential. For the particular examples of one- and
three-dimensional harmonic traps we explore the microcanonical statistics
in detail, with the help of the saddle-point method. Emphasizing the close
connection between the partition theory of integer numbers and the statistical
mechanics of ideal Bosons in one-dimensional harmonic traps, and utilizing
thermodynamical arguments, we also derive an accurate formula for the
fluctuation of the number of summands that occur when a large integer
is partitioned.
\end{abstract}


\hfill
\begin{minipage}{73mm}
{\em There is, essentially, only one problem in statistical thermodynamics:
the distribution of a given amount of energy~$E$ over $N$~identical systems.}

\vspace{3mm}

Erwin Schr\"odinger~\cite{Schrodinger46}

\vspace{10mm}
\end{minipage}

\section{Introduction}
\label{S1}

The ideal Bose gas is customarily treated in the grand canonical ensemble,
since the evaluation of the canonical partition sum is impeded by the
constraint that the total particle number $N$ be fixed. In contrast, after
introducing a variable that is conjugate to $N$, the fugacity $z$, the
computation of the ensuing grand canonical partition function $\Xi(z,\beta)$
requires merely the summation of geometric series, and all thermodynamic
properties of the Bose gas are then obtained by taking suitable derivatives
of $\Xi(z,\beta)$ with respect to $z$ or the inverse temperature $\beta$. 

There is, however, one serious failure of the grand canonical ensemble. 
Grand canonical statistics predicts that the mean-square fluctuation
$\langle \delta^2 n_{\nu} \rangle_{gc}$ of the $\nu$-th single-particle
level's occupation equals
$\langle n_\nu \rangle_{gc} \, (\langle n_\nu \rangle_{gc} + 1)$.
Applied to the ground state $\nu = 0$, this gives
\[
\langle \delta^2 n_{0} \rangle_{gc} \; = \;
    \langle n_0 \rangle_{gc} \, \left( \langle n_0 \rangle_{gc} + 1 \right)
\]
even when the temperature $T$ approaches zero, so that all $N$ particles
condense into the ground state. But the implication of huge fluctuations,
$\langle \delta^2 n_{0} \rangle_{gc} = N(N+1)$, is clearly unacceptable;
when all particles occupy the ground state, the fluctuation has to die out.

This grand canonical fluctuation catastrophe has been discussed by generations
of physicists, and possible remedies have been suggested within the canonical
framework~\cite{Fierz56,FujiwaraEtAl70,ZiffEtAl77}. After the recent
realization of Bose--Einstein condensates of weakly interacting gases of
alkali atoms~\cite{Anderson95,Davis95,Bradley97,Ernst98,MiesnerKetterle98}
had brought the problem back into the focus of
interest~\cite{GH96,Politzer96,GajdaRzazewski97,WilkensWeiss97},    
significant steps towards its general solution could be made. Ideal 
condensate fluctations have been computed for certain classes of
single-particle spectra, i.e., for certain trap types, both within the
canonical ensemble~\cite{Politzer96,WeissWilkens97}, where the gas is
still exchanging energy with some hypothetical heat bath, and within the
more appropriate microcanonical ensemble~\cite{GH96,NavezEtAl97,GH97b},
where it is completely isolated. Interestingly, canonical and microcanonical
fluctuations have been found to agree in the large-$N$-limit for
one-dimensional harmonic trapping potentials~\cite{GH96,WilkensWeiss97},
but to differ in the case of three-dimensional isotropic harmonic
traps~\cite{Politzer96,NavezEtAl97}.

Yet, in the true spirit of theoretical physics one would clearly like
to have more than merely some formulas for condensate fluctuations in
particular traps. Can't one extract a common feature that underlies those
formulas, such that simply inspecting that very feature allows one to
determine, without any actual calculation, the temperature dependence of
the condensate fluctuation, and to decide whether or not canonical and
microcanonical fluctuations are asymptotically equal?

It is such a refined understanding that we aim at in the present work.
As will be shown, the feature imagined above actually exists: It is the
rightmost pole, in the complex $t$-plane, of the Zeta function 
\[
Z(\beta,t) \; = \;
    \sum_{\nu=1}^{\infty} \frac{1}{(\beta\varepsilon_\nu)^t}
\]
that is furnished by the system's single-particle energies $\varepsilon_\nu$.
In order to substantiate this statement, we will proceed as follows:
We start in the next section by deriving simple expressions that relate
the canonical number of ground state particles $\langle n_0 \rangle_{cn}$, 
and its mean-square fluctuation $\langle \delta^2 n_0 \rangle_{cn}$, to
$Z(\beta,t)$. The key point exploited there is the approximate equivalence
of a trapped Bose gas in the condensate regime to a system of Boltzmannian
harmonic oscillators. This equivalence, which holds irrespective of the
particular form of the trapping potential, implies that both
$\langle n_0 \rangle_{cn}$ and $\langle \delta^2 n_0 \rangle_{cn}$
can be expressed in terms of harmonic oscillator sums, which explain the
emergence of the spectral Zeta function $Z(\beta,t)$ and can be computed
with the help of well-established techniques~\cite{KirstenToms96a}.
In Section~\ref{S3} we then evaluate these general canonical formulas for
$d$-dimensional isotropic harmonic traps, where $Z(\beta,t)$ reduces to
ordinary Riemann Zeta functions, and for anisotropic harmonic traps,
where it leads to Zeta functions of the Barnes type.   
  
In Section~\ref{S4} we compare the canonical statistics of harmonically
trapped gases for $d = 1$ and $d = 3$ to their microcanonical counterparts.
The strategy adopted there --- the calculation of microcanonical moments
from the easily accessible corresponding canonical moments by means of
saddle-point inversions --- is technically rather cumbersome and certainly
not to be recommended if one merely wishes to obtain the microcanonical
condensate fluctuations $\langle \delta^2 n_0 \rangle_{mc}$, but
it explains in precise detail just how the difference between the
canonical and the microcanonical ensemble comes into play, and why
$\langle \delta^2 n_0 \rangle_{mc}$ becomes asymptotically equal to
$\langle \delta^2 n_0 \rangle_{cn}$ in some cases, but not in others.
A convenient expression for the immediate determination of
$\langle \delta^2 n_0 \rangle_{cn} - \langle \delta^2 n_0 \rangle_{mc}$,
based again on the spectral Zeta function $Z(\beta,t)$, is then derived
in Section~\ref{S5}. The final Section~\ref{S6} contains a concluding
discussion; three appendices offer technical details.

Since we restrict ourselves to non-interacting Bose gases, the main value of
the present work lies on the conceptual side --- after all, the ideal Bose
gas ought to be properly understood ---, but it may well turn out to be of
more than purely academical importance: After it has been demonstrated now
that the $s$-wave scattering length in optically confined condensates can be
tuned by varying an external magnetic field~\cite{Inouye98}, the creation of
almost ideal Bose--Einstein condensates might become feasible. If it did,
then also the experimental investigation of the basic statistical questions
studied here, such as the connection between the temperature dependence
of the condensate fluctuation and the properties of the trap potential,
should not remain out of reach.

Finally, there is still another appealing side-aspect: Since the
microcanonical statistics of ideal Bosons in one-dimensional harmonic traps
can be mapped to the partition theory of integer numbers, a natural by-product
of our work is a fairly accurate formula for the fluctuation of the number
of integer parts into which a large integer may by decomposed. The derivation
of that formula in Section~\ref{S5} is a beautiful example for the deep-rooted
connection between partition theory and statistical mechanics, quite in the
sense of Schr\"odinger's remark quoted above.

\section{Canonical description of ideal Bose--Einstein condensates}
\label{S2}

We consider a gas of non-interacting Bose particles confined in a trap with
discrete single-particle energies $\varepsilon_{\nu}$ ($\nu = 0,1,2,\ldots$)
and stipulate that the ground state energy be equal to zero,
$\varepsilon_0 = 0$. Starting from the grand canonical partition sum
\begin{equation}
\prod_{\nu=0}^{\infty} \frac{1}{1 - z\exp(-\beta\varepsilon_{\nu})}
    \; = \; \Xi(z,\beta)  \; ,
\label{GCPF}
\end{equation}
where $\beta = 1/(k_B T)$ is the inverse temperature, we have the
familiar expansion
\begin{equation}
\Xi(z,\beta) \; = \; \sum_{N=0}^{\infty} z^N \sum_{E} e^{-\beta E} \, 
    \Omega(E|N)	\; ,
\end{equation}
with coefficients $\Omega(E|N)$ denoting the number of microstates
accessible to an $N$-particle gas with total excitation energy $E$.
Combinatorically speaking, $\Omega(E|N)$ is the number of possibilities
for sharing the energy $E$ among {\em up to\/} $N$ particles --- the
number $N_{\rm ex}$ of particles that are actually excited and thus carry
a part of $E$ remains unspecified.

The clear distinction between $N$ and $N_{\rm ex}$ is the starting point
for studying statistical properties of Bose--Einstein condensates in gases
with fixed particle number. When $N_{\rm ex}$ out of $N$ Bose particles
are excited, there remain $N - N_{\rm ex}$ particles forming the
condensate, and the corresponding number of microstates (that is, the
number of possibilities for distributing the excitation energy $E$ over
{\em exactly\/} $N_{\rm ex}$ particles) is given by
\begin{equation}
\Omega(E|N_{\rm ex}) - \Omega(E|N_{\rm ex}-1) \; \equiv \;
    \Phi(N_{\rm ex}|E) \; .
\end{equation}
Within the {\em canonical\/} ensemble, i.e., if the $N$-particle gas is
in contact with some heat bath of temperature $T$, the probability for
finding $N_{\rm ex}$ excited particles can then be written as
\begin{equation}
p_{cn}(N_{\rm ex},\beta) \; = \;
    \frac{\sum_E e^{-\beta E} \, \Phi(N_{\rm ex}|E)}
    {\sum_E e^{-\beta E}
    \sum_{N^{'}_{\rm ex}=0}^{N} \Phi(N^{'}_{\rm ex}|E)}
    \quad , \quad N_{\rm ex} \le N \; .
\label{CNDI}
\end{equation}
The expectation value $\langle N_{\rm ex} \rangle_{cn}$ with respect to
this distribution yields the canonical ground state occupation number,
\begin{equation}
\langle n_0 \rangle_{cn} \; = \; N - \langle N_{\rm ex} \rangle_{cn} \; ;
\end{equation}
the canonical mean-square fluctuation of the number of condensate particles
is identical to the fluctuation of the number of excited particles,
\begin{eqnarray}
\langle \delta^2 n_0 \rangle_{cn} & = &
    \langle \delta^2 N_{\rm ex} \rangle_{cn}	\nonumber   \\
    & = &
    \langle N_{\rm ex}^2 \rangle_{cn}
  - \langle N_{\rm ex} \rangle_{cn}^2	\; .
\end{eqnarray}
  
In order to calculate these cumulants, we consider the function    
\begin{equation}
(1 - z)\, \Xi(z,\beta) \; \equiv \; \Xi_{\rm ex}(z,\beta)    \; ,
\end{equation}
which satisfies the identities
\begin{eqnarray}
\Xi_{\rm ex}(z,\beta) & = & \sum_{N=0}^{\infty} \left(z^N - z^{N+1}\right)
    \sum_{E} e^{-\beta E} \, \Omega(E|N)	\nonumber   \\
    & = & \sum_{N=0}^{\infty} z^N
    \sum_{E} e^{-\beta E} \left[ \Omega(E|N) - \Omega(E|N-1) \right] 
\end{eqnarray}
with $\Omega(E|-1) = 0$. Hence, replacing the summation index $N$ by
$N_{\rm ex}$, one finds  
\begin{equation}
\Xi_{\rm ex}(z,\beta) \; = \; \sum_{N_{\rm ex}=0}^{\infty} z^{N_{\rm ex}}
    \sum_E e^{-\beta E} \, \Phi(N_{\rm ex}|E)  \; . 
\end{equation}
On the other hand we have
\begin{equation}
\Xi_{\rm ex}(z,\beta) \; = \; 
    \prod_{\nu=1}^{\infty} \frac{1}{1 - z\exp(-\beta\varepsilon_{\nu})} \; ,
\label{GRPF}
\end{equation}
where, in contrast to Eq.~(\ref{GCPF}), the product runs only over the
excited states $\nu \geq 1$: {\em The grand canonical partition sum of
a fictituous Bose gas which emerges from the actual gas by removing the
single-particle ground state is the generating function for
$\Phi(N_{\rm ex}|E)$}. Differentiating this generating function $k$~times,
and then setting $z = 1$, one gets the canonical moments~\footnote{This
    corresponds to a trick suggested by D.\ H.\ Lehmer and taken up
    by L.\ B.\ Richmond for calculating $k$-th moments of partitions
    of integer numbers; see Ref.~\cite{Richmond75}. Note that the
    asymptotic moment formula derived in that paper is not correct for
    $k \ge 2$; the corrected formula is stated in Appendix~\ref{AC}.}
\begin{equation}
\left.
\left(z\frac{\partial}{\partial z}\right)^{\!k} \!
    \Xi_{\rm ex}(z,\beta) \right|_{z=1} = \;
    \sum_E e^{-\beta E} \left(
    \sum_{N_{\rm ex}=0}^{\infty} N_{\rm ex}^k \, \Phi(N_{\rm ex}|E) \right)
    \; \equiv \; M_k(\beta) \; .
\label{CNKM}
\end{equation}
In the customary grand canonical framework, $z$ is identified with the
fugacity and linked to the ground state occupation number
$\langle n_0 \rangle_{gc}$ by
$z = (1 + 1/\langle n_0 \rangle_{gc})^{-1}$. In that case $z$ remains
strictly less than unity, thus preventing the ground state factor
in Eq.~(\ref{GCPF}) from diverging. In contrast, $z$ is no more than a
formal parameter in the present analysis, entirely unrelated to the ground
state occupation; and since the ground state factor is absent in the
generating function $\Xi_{\rm ex}(z,\beta)$, there is nothing to prevent
us from fixing $z = 1$.
  
If the sum over $N_{\rm ex}$ in Eq.~(\ref{CNKM}) did not range from
$0$ to $\infty$, but instead from $0$ to the actual particle number $N$, as
it does in the canonical distribution~(\ref{CNDI}), then the ratio
$M_1(\beta)/M_0(\beta)$ would be {\em exactly\/} equal to the canonical
expectation value $\langle N_{\rm ex}\rangle_{cn}$. But even if we do not
have an exact equality here, the difference between these two quantities must
be negligible {\em if there is a condensate\/}. This statement requires
no proof, it is a mere tautology: a condensate can only be present if
those microstates where the energy $E$ is spread over all $N$ particles
are statistically negligible, so that also the microstates that would become
available if additional zero-energy (ground state) particles were added to the
gas cannot make themselves felt. Hence, in the presence of a Bose--Einstein
condensate we have, for small~$k$,  
\begin{equation}
\sum_{N_{\rm ex}=0}^{\infty} N_{\rm ex}^k \, \Phi(N_{\rm ex}|E) \; = \;
\sum_{N_{\rm ex}=0}^{N} N_{\rm ex}^k \, \Phi(N_{\rm ex}|E)
\label{OSAP}
\end{equation}
at least to a very good approximation, which gives both 
\begin{equation}
\langle N_{\rm ex} \rangle_{cn} \; = \; \frac{M_1(\beta)}{M_0(\beta)} 
\label{CNGR}
\end{equation}
and
\begin{equation}
\langle \delta^2 N_{\rm ex} \rangle_{cn} \; = \; \frac{M_2(\beta)}{M_0(\beta)}
    - \left( \frac{M_1(\beta)}{M_0(\beta)} \right)^2 \; .
\label{CNFL}
\end{equation}

The approximation~(\ref{OSAP}), expressing the replacement of the actual
condensate of $N - \langle N_{\rm ex} \rangle_{cn}$ particles by a
condensate consisting of infinitely many particles, can be interpreted in two
different ways. On the one hand, the infinitely many ground state particles
may be regarded as forming a particle reservoir for the excited-states
subsystem. Such an approach to computing canonical condensate fluctuations
had been suggested as early as 1956 by Fierz~\cite{Fierz56}; it corresponds
to the ``Maxwell's Demon Ensemble'' recently put forward by
Navez et al.~\cite{NavezEtAl97}.

The second interpretation rests on the observation that for $k = 0$
the approximation~(\ref{OSAP}) takes the form
\begin{equation}
\sum_{N_{\rm ex}=0}^{\infty} \Phi(N_{\rm ex}|E) \; = \; \Omega(E|N) \; , 
\label{ENTR}
\end{equation}
so that Eq.~(\ref{CNKM}) gives
\begin{equation}
\sum_E e^{-\beta E} \, \Omega(E|N) \; = \;
    \prod_{\nu = 1}^{\infty} \frac{1}{1 - \exp(-\beta\varepsilon_\nu)} \;
    \; \equiv \; Z(\beta) \; . 
\label{BOSY}
\end{equation}
Since each factor $1/[1 - \exp(-\beta\varepsilon_\nu)]$ corresponds to
a geometric series, i.e., to the canonical partition function of a
simple harmonic oscillator with frequency $\varepsilon_\nu/\hbar$,
Eq.~(\ref{BOSY}) states that if there is a condensate, so that
Eq.~(\ref{ENTR}) holds, then the canonical partition function of an ideal
Bose gas with arbitrary single-particle energies is well approximated by the
canonical partition function of a system of distinguishable harmonic
oscillators, each excited single-particle level $\varepsilon_{\nu}$
corresponding to an oscillator with frequency $\varepsilon_{\nu}/\hbar$.
Thus, {\em for temperatures below the onset of Bose--Einstein condensation
the thermodynamics of the actual Bose gas practically coincides with the
thermodynamics of a Boltzmannian harmonic oscillator system\/}, regardless
of the specific form of the trapping potential. For this reason, we will
refer to the approximation~(\ref{OSAP}) as the {\em oscillator
approximation\/}. For the particular case of a three-dimensional isotropic
harmonic trapping potential, the quality of this approximation has been
confirmed in Ref.~\cite{GH97a} by comparing the entropy of the actual
Bose gas with that of its Boltzmannian substitute.

Within this oscillator approximation, the determination of the number
$\langle N_{\rm ex} \rangle_{cn}$ of excited particles, and of the   
canonical mean-square condensate fluctuation
$\langle \delta^2 n_0 \rangle_{cn} = \langle \delta^2 N_{\rm ex} \rangle_{cn}$,
becomes remarkably simple. Doing the derivatives demanded by Eq.~(\ref{CNKM}),
we find
\begin{eqnarray}
M_0(\beta) & = & Z(\beta)
\label{CN0M}	\\
M_1(\beta) & = & Z(\beta) S_1(\beta)
\\
M_2(\beta) & = & Z(\beta) \left[ S_1^2(\beta) + S_2(\beta) \right]  \; ,
\label{CN2M}
\end{eqnarray}
with $Z(\beta)$ as given by Eq.~(\ref{BOSY}), and
\begin{eqnarray}
S_1(\beta) & = & \sum_{\nu=1}^{\infty}
    \frac{1}{\exp(\beta\varepsilon_\nu) - 1}
    \nonumber	\\
    & = & \sum_{\nu=1}^{\infty} \, \sum_{r=0}^{\infty} \,
    \exp[-\beta\varepsilon_\nu(r+1)]	\; , 	 
    \label{SUM1}    \\
S_2(\beta) & = & \sum_{\nu=1}^{\infty}
    \frac{1}{\exp(\beta\varepsilon_\nu) - 1}
    \left( \frac{1}{\exp(\beta\varepsilon_\nu) - 1} + 1 \right)
    \nonumber	\\
    & = & \sum_{\nu=1}^{\infty} \, \sum_{r=1}^{\infty} \,
    r \, \exp[-\beta\varepsilon_\nu r] \; .
    \label{SUM2}
\end{eqnarray}
Computing the ratios $M_1(\beta)/M_0(\beta)$ and $M_2(\beta)/M_0(\beta)$
according to Eqs.~(\ref{CNGR}) and (\ref{CNFL}), the oscillator partition
function $Z(\beta)$ drops out (and, hence, does not even have to be evaluated
here!), and we arrive at the appealing relations
\begin{eqnarray}
\langle N_{\rm ex} \rangle_{cn} & = & S_1(\beta)
\label{NEXS} \\
\langle \delta^2 N_{\rm ex} \rangle_{cn} & = & S_2(\beta) \; .
\label{FLUS}
\end{eqnarray}
Since $1/[\exp(\beta\varepsilon_\nu) - 1] = \langle n_\nu \rangle_{gc}$
is just the grand canonical expectation value for the occupation of the
$\nu$-th excited state in a partially condensed Bose gas (the fugacity
of which is $z = 1$), and
$\langle n_\nu \rangle_{gc} \, (\langle n_\nu \rangle_{gc} + 1)
= \langle \delta^2 n_{\nu} \rangle_{gc}$ is the corresponding
grand canonical mean-square fluctuation, the respresentations (\ref{SUM1})
and (\ref{SUM2}) reveal that the canonical expectation value of the number
of excited particles equals the grand canonical one, and that the canonical
mean-square fluctuation of the ground state occupation number can simply be
computed by adding the grand canonical fluctuations of the excited levels,
subject to only the oscillator approximation. This is precisely what had
been anticipated by Fierz~\cite{Fierz56}, and what has also been exploited
in a heuristic manner by Politzer~\cite{Politzer96} when investigating 
the three-dimensional harmonic trap.     

For evaluating the sums $S_1(\beta)$ and $S_2(\beta)$ we employ the
Mellin--Barnes integral representation~\cite{KirstenToms96a}
\begin{equation}
e^{-\alpha} \; = \; \frac{1}{2\pi i}\int_{\tau-i\infty}^{\tau+i\infty}
    \! {\mbox d} t \, \alpha^{-t} \, \Gamma(t)   \; ,
\end{equation}
valid for real $\tau > 0$ and complex $\alpha$ with $R\!e\,(\alpha) > 0$.
This leads to
\begin{eqnarray}
\langle N_{\rm ex} \rangle_{cn} & = &
    \sum_{\nu=1}^{\infty} \, \sum_{r=0}^{\infty}
    \frac{1}{2\pi i}\int_{\tau-i\infty}^{\tau+i\infty} \! {\mbox d} t \,
    \frac{\Gamma(t)}{[\beta\varepsilon_\nu(r+1)]^t}	\nonumber \\
    & = & 
    \frac{1}{2\pi i}\int_{\tau-i\infty}^{\tau+i\infty} \! {\mbox d} t \,
    \sum_{\nu=1}^{\infty} \, \sum_{r=0}^{\infty}
    \frac{\Gamma(t)}{[\beta\varepsilon_\nu(r+1)]^t}     \nonumber \\
    & = &
    \frac{1}{2\pi i}\int_{\tau-i\infty}^{\tau+i\infty} \! {\mbox d} t \,
    \Gamma(t) Z(\beta,t) \zeta(t)   \; ,
\label{MBGR}
\end{eqnarray}
where $\zeta(t) = \sum_{r=1}^{\infty} r^{-t}$ denotes the Riemann Zeta
function, and we have introduced the spectral Zeta function
\begin{equation}
Z(\beta,t) \; = \;
    \sum_{\nu=1}^{\infty} \frac{1}{(\beta\varepsilon_\nu)^t}
\label{SZFN}
\end{equation}
that embodies the necessary information about the trap spectrum. In the same
way we find the remarkably similar-looking equation
\begin{equation}
\langle \delta^2 N_{\rm ex} \rangle_{cn} \; = \; 
    \frac{1}{2\pi i}\int_{\tau-i\infty}^{\tau+i\infty} \! {\mbox d} t \,
    \Gamma(t) Z(\beta,t) \zeta(t-1)   \; .
\label{MBFL}
\end{equation}
It should be noted that interchanging summations and integration
in Eq.~(\ref{MBGR}), and in the analogous derivation of the canonical
fluctuation formula~(\ref{MBFL}), requires the absolute convergence of the
emerging sums. Therefore, the real number $\tau$ has to be chosen such that
the path of integration up the complex $t$-plane lies to the right of the
poles of both Zeta functions. 

Now the temperature dependence of $\langle N_{\rm ex} \rangle_{cn}$
or $\langle \delta^2 N_{\rm ex} \rangle_{cn}$ is determined by the pole
of the integrand~(\ref{MBGR}) or (\ref{MBFL}) that lies farthest to the
right. Since $\Gamma(t)$ has poles merely at $t = 0, -1, -2, \ldots\,$,
the decisive pole is provided {\em either\/} by the Riemann Zeta function
$\zeta(t)$ or $\zeta(t-1)$, respectively, {\em or\/} by its spectral opponent
$Z(\beta,t)$, which depends on the particular trap under study~\cite{GH97b}.
This competition will be discussed in detail in the following section,
focussing on harmonic trapping potentials.

\section{Isotropic and anisotropic harmonic traps} 
\label{S3}

The evaluation of the canonical relations~(\ref{MBGR}) and (\ref{MBFL})
reduces to a mere formality if the pole structure of the spectral Zeta
function~(\ref{SZFN}) is known. The simplest examples are provided by
$d$-dimensional isotropic harmonic traps, since then $Z(\beta,t)$ becomes
a sum of Riemannian Zeta functions. Namely, denoting the angular frequency
of such a trap by $\omega$, the degree of degeneracy $g_\nu$ of a
single-particle state with excitation energy $\nu\hbar\omega$ is
\begin{equation}
g_\nu \; = \; \left( \begin{array}{c}
    \nu + d-1 \\ d-1 \end{array} \right)    \; , 
\end{equation}
so that $Z(\beta,t)$ acquires the form
\begin{equation}
Z(\beta,t) \; = \; (\beta\hbar\omega)^{-t} \sum_{\nu = 1}^{\infty} \, 
    \frac{g_{\nu}}{\nu^t}   \; ,
\end{equation}
giving in explicit terms
\begin{equation}
\begin{array}{ccll}
Z(\beta,t) & = & (\beta\hbar\omega)^{-t} \zeta(t)
    & \quad {\mbox{for}} \; d = 1 \; ,   \\
Z(\beta,t) & = & (\beta\hbar\omega)^{-t} \left[ \zeta(t-1) + \zeta(t) \right]
    & \quad {\mbox{for}} \; d = 2 \; ,   \\
Z(\beta,t) & = & (\beta\hbar\omega)^{-t} \left[ \zeta(t-2)/2
    + 3\zeta(t-1)/2 + \zeta(t) \right]
    & \quad {\mbox{for}} \; d = 3 \; .
\end{array}
\end{equation}
We now aim at the temperature dependence of $\langle N_{\rm ex} \rangle_{cn}$
and $\langle \delta^2 N_{\rm ex} \rangle_{cn}$ for temperatures below the
onset of a ``macroscopic'' ground state occupation, so that the oscillator
approximation retains its validity, but well above the level spacing
temperature $\hbar\omega/k_B$, so that $\beta\hbar\omega \ll 1$. Such a
temperature interval exists if the particle number $N$ is sufficiently large,
since the condensation temperature generally increases with $N$. The desired
asymptotic $T$-dependence can then directly be read off from the residue of
the rightmost pole of the respective integrand~(\ref{MBGR}) or (\ref{MBFL}).
Since $\zeta(z)$ has merely a single pole at $z = 1$, simple and with residue
$+1$, namely~\cite{WhittakerWatson62}
\begin{equation}
\zeta(z) \; \approx \; \frac{1}{z-1} + \gamma
\label{riemann}
\end{equation}
for $z$ close to $1$, the calculation of that residue is particularly easy
if the rightmost pole in~(\ref{MBGR}) or (\ref{MBFL}) is simple. In the case
of a double pole we also need the identity
\begin{eqnarray}
\Gamma'(n) & = & \Gamma(n) \, \psi(n)   \nonumber	\\
    & = & \Gamma(n) \left(-\gamma
        + \sum_{m=1}^{n-1} \frac{1}{m} \right) 
\end{eqnarray}
for the Psi function at integer arguments, with $\gamma \approx 0.5772$
denoting Euler's constant. This is the only technical knowledge required
for computing the number of excited particles, and its fluctuation,
in a $d$-dimensional isotropic harmonic trap within the canonical ensemble:
\begin{description}
\item[(i)] For $d = 1$, the number of excited particles is governed
    by the double pole at $t = 1$ which emerges since $Z(\beta,t)$ now
    is proportional to $\zeta(t)$, whereas the mean-square fluctuation
    is dominated by the simple pole of $\zeta(t-1)$ at $t = 2$:
    \begin{eqnarray}
    \langle N_{\rm ex} \rangle_{cn}		& = &
    \frac{k_B T}{\hbar\omega}
    \left[ \ln \! \left(\frac{k_B T}{\hbar\omega}\right) + \gamma \right]
    \label{GRD1}    \\ 
    \langle \delta^2 N_{\rm ex} \rangle_{cn}    & = &
    \left(\frac{k_B T}{\hbar\omega}\right)^2 \zeta(2) \; .
    \label{FLD1}	
    \end{eqnarray}
\item[(ii)] For $d = 2$, the rightmost pole of $Z(\beta,t)$ has moved to
    $t = 2$ and thus determines $\langle N_{\rm ex} \rangle_{cn}$ all
    by itself, but now the product $Z(\beta,t) \zeta(t-1)$ provides
    a double pole that governs the asymptotics of the fluctuation:
    \begin{eqnarray}
    \langle N_{\rm ex} \rangle_{cn}           & = &
    \left(\frac{k_B T}{\hbar\omega}\right)^2 \zeta(2)
    \label{GRD2}    \\
    \langle \delta^2 N_{\rm ex} \rangle_{cn}  & = &
    \left(\frac{k_B T}{\hbar\omega}\right)^2
    \left[ \ln \! \left(\frac{k_B T}{\hbar\omega}\right)
	+ \gamma + 1 + \zeta(2) \right] \; .
    \label{FLD2}
    \end{eqnarray}
\item[(iii)] For $d = 3$, the pole of the spectral Zeta function $Z(\beta,t)$
    at $t = 3$ wins in both cases:
    \begin{eqnarray}
    \langle N_{\rm ex} \rangle_{cn}          & = &
    \left(\frac{k_B T}{\hbar\omega}\right)^3 \zeta(3)
    \label{GRD3}    \\
    \langle \delta^2 N_{\rm ex} \rangle_{cn} & = &
    \left(\frac{k_B T}{\hbar\omega}\right)^3 \zeta(2)	\; .
    \label{FLD3}
    \end{eqnarray}
\end{description}
Of course, these results remain valid only as long as
$\langle N_{\rm ex} \rangle_{cn} < N$.
Equating $\langle N_{\rm ex} \rangle_{cn}$ and $N$ for $d = 3$, e.g.,
one finds the large-$N$ condensation temperature
\begin{equation}
T_0 \; = \; \frac{\hbar\omega}{k_B}\left(\frac{N}{\zeta(3)}\right)^{1/3}
\end{equation}
for an ideal gas in a three-dimensional harmonic trap that exchanges energy,
but no particles, with a heat bath. As expected, this result agrees with the
one provided by the familiar grand canonical ensemble~\cite{deGrootEtAl50}.
Even more, taking into account also the next-to-leading pole, one obtains
the improvement
\begin{equation}
\langle N_{\rm ex} \rangle_{cn} \; = \;
    \zeta(3) \left(\frac{k_B T}{\hbar\omega}\right)^3  
  + \frac{3}{2} \, \zeta(2) \left(\frac{k_B T}{\hbar\omega}\right)^2
\end{equation}
to Eq.~(\ref{GRD3}), implying that for Bose gases with merely a moderate
number of particles the actual condensation temperature $T_C$ is lowered by
terms of the order $N^{-1/3}$ against $T_0$,  
\begin{equation}
T_C \; = \; T_0 \left[1 - \frac{\zeta(2)}{2\,\zeta(3)^{2/3}}
    \frac{1}{N^{1/3}}\right] \; .
\end{equation}
Even this improved canonical expression equals its grand canonical
counterpart~\cite{GH95,KetterleDruten96,KirstenToms96b,HaugerudEtAl97}.

These examples nicely illustrate the working principle of the basic
integral representations~(\ref{MBGR}) and (\ref{MBFL}): There are two
opponents that place poles on the positive real axis, namely the spectral
Zeta function $Z(\beta,t)$ on the one hand, which depends on the particular
trap, and $\zeta(t)$ or $\zeta(t-1)$ on the other, which are entirely
independent of the system. For both the number of excited particles and its
mean-square fluctuation, the exponent of $T$ is given by the location of the
pole farthest to the right. Where\-as the pole of $\zeta(t)$ and $\zeta(t-1)$
does, naturally, not depend on the spatial dimension~$d$, the asymptotically
relevant pole of $Z(\beta,t)$ lies at $t = d$ and thus moves with increasing
$d$ to the right, governing $\langle N_{\rm ex} \rangle_{cn}$ above $d = 1$ and
$\langle \delta^2 N_{\rm ex} \rangle_{cn}$ above $d = 2$.  

But what about the {\em anisotropic\/} traps that play a major role in
present experiments? With the ground state energy set to zero, and angular
trap frequencies $\omega_i$ ($i = 1,\dots,d$), the energy levels then are
\begin{equation}
\varepsilon_{\nu_1,\ldots,\nu_d} \; = \;
    \hbar (\omega_1 \nu_1 + \ldots + \omega_d \nu_d ) 
    \; \equiv \; \hbar \vec \omega \vec \nu \; ,
    \qquad \vec \nu \in \nats_0^d \; .
\label{enani}
\end{equation}
The spectral Zeta function 
\begin{equation}
Z(\beta, t) \; =  \sum_{\vec \nu \in \nats_0^d / \{0\}} 
    \frac{1}{(\beta\hbar \vec \omega \vec \nu )^t} 
\label{specani}
\end{equation}
now is a Zeta function of the Barnes type~\cite{Barnes03} (see also 
Ref.~\cite{Dowker94}). Its rightmost pole is located at $t=d$, with residue 
\begin{equation}
\mbox{res} \; Z(\beta, d) \; = \; \frac 1 {\Gamma (d)}
    \left(\frac {k_B T}{\hbar \Omega}\right)^d \; ,
\label{resbarnes}
\end{equation}
where we have introduced the geometric mean $\Omega$ of the trap frequencies, 
\begin{equation}
\Omega \; = \; \left( \prod_{i=1}^d \omega_i \right)^{1/d}
\label{geomean}
\end{equation}
The derivation of Eq.~(\ref{resbarnes}) is sketched in Appendix~\ref{AA}.
 
The asymptotic evaluation of the canonical formulas~(\ref{MBGR})
and~(\ref{MBFL}) now requires $\beta\hbar\omega_i \ll 1$ for all~$i$.
If this condition is not met, since, for instance, one of the trapping
frequencies is much larger than the others, one has to treat the entailing
dimensional crossover effects~\cite{DrutenKetterle97} by keeping the
corresponding part of $Z(\beta,t)$ as a discrete sum. In the following we
will assume merely moderate anisotropy, so that the above inequalities are
satisfied.   

For two-dimensional anisotropic harmonic traps, the computation of the
canonically expected number of excited particles, and its fluctuation,
then leads to
\begin{eqnarray}
\langle N_{\rm ex} \rangle_{cn} & = &
    \left(\frac {k_B T}{\hbar \Omega }\right)^2 \zeta (2)
    \label{ani2}    \\
\langle \delta^2 N_{\rm ex} \rangle_{cn} & = &
    \left(\frac {k_B T}{\hbar \Omega }\right)^2
    \left[\ln\!\left( \frac{k_B T}{\hbar (\omega_1 + \omega_2)}\right)
  + \left(\frac{\omega_1}{\omega_2} + \frac{\omega_2}{\omega_1} \right)
    \zeta (2) + I(\omega_1,\omega_2) \right] \; ,
    \label{delani2}
\end{eqnarray}
with 
\begin{equation}
I(\omega_1,\omega_2) \; = \; \int_0^{\infty} \! {\mbox d} \alpha \, 
    \alpha e^{-\left(\sqrt{\frac{\omega_1}{\omega_2}} + 
                \sqrt{\frac{\omega_2}{\omega_1}}\right) \alpha } 
    \left(\frac{1}{\left(1-e^{-\sqrt{\frac{\omega_1}{\omega_2}}\alpha}\right)
                   \left(1-e^{-\sqrt{\frac{\omega_2}{\omega_1}}\alpha}\right)}  
   -\frac{1}{\alpha^2} \right) \; .
\label{finiteb}
\end{equation}

Equation~(\ref{delani2}) reveals a rather complicated dependence of 
the fluctuation $\langle \delta^2 N_{\rm ex}\rangle_{cn}$ on the trap 
frequencies $\omega_1$ and $\omega_2$. The comparatively simple form
of the previous Eq.~(\ref{FLD2}) for an isotropic trap has its technical
reason in the simple expansion~(\ref{riemann}) of $\zeta(z)$ around its pole.
In contrast, for two-dimensional anisotropic traps we need the analogous
expansion of the Barnes Zeta function~(\ref{specani}) for $d = 2$. 
The finite part of this expansion, corresponding to Euler's constant 
$\gamma$ in Eq.~(\ref{riemann}), now becomes a function of the frequencies 
$\omega_1$ and $\omega_2$ that enters into the above result. Details are
explained in Appendix~\ref{AB}, where we also show the identity 
\begin{equation}
I(\omega,\omega) \; = \; \gamma + 1 + \ln 2 - \zeta (2)	    \; , 
\label{isolimit} 
\end{equation}
which ensures that Eq.~(\ref{delani2}) reduces to the isotropic
result~(\ref{FLD2}) for $\omega_1 = \omega_2 = \omega$.

For any dimension $d \geq 3$, it is the pole of $Z (\beta, t)$ at $t = d$
which determines the behaviour of both $\langle N_{\rm ex} \rangle _{cn}$
and $\langle \delta^2 N_{\rm ex} \rangle_{cn}$:
\begin{eqnarray}
\langle N_{\rm ex} \rangle _{cn} & = & 
    \left(\frac {k_B T}{\hbar \Omega}\right)^d \zeta (d)
    \label{anid}    \\
\langle \delta^2 N_{\rm ex} \rangle_{cn} & = &
    \left(\frac {k_B T}{\hbar \Omega }\right)^d \zeta (d-1)   \; .
    \label{delanid}
\end{eqnarray}
The difference between the isotropic and the mildly anisotropic case now
merely consists in the replacement of the frequency $\omega$ by the
geometric mean $\Omega$.

\section{A saddle-point approach to microcanonical statistics}
\label{S4}

When the ideal $N$-particle Bose gas is completely isolated from its
surrounding, carrying a total excitation energy $E$, one has to resort
to the microcanonical framework. The micro\-canonical counterpart of the
distribution~(\ref{CNDI}), that is, the probability for finding
$N_{\rm ex}$ out of the $N$ isolated particles in an excited state, is
given by 
\begin{equation}
p_{mc}(N_{\rm ex},E) \; = \;
    \frac{\Phi(N_{\rm ex}|E)}
         {\sum_{N^{'}_{\rm ex}=0}^{N} \Phi(N^{'}_{\rm ex}|E)}
    \quad , \quad N_{\rm ex} \le N \; .
\label{MCDI}
\end{equation}
It is quite instructive to copy the previous canonical analysis as far
as possible, in order to pin down precisely how the difference between
the canonical and the microcanonical ensemble manifests itself. Hence, we
wish to calculate the $k$-th moments
\begin{equation}
\mu_k(E) \; = \; \sum_{N_{\rm ex}=0}^{N} N_{\rm ex}^k \, \Phi(N_{\rm ex}|E)	
\label{MIMO}
\end{equation}
of this distribution~(\ref{MCDI}), which yield the microcanonical
expectation value
\begin{equation}
\langle N_{\rm ex} \rangle_{mc} \; = \; \frac{\mu_1(E)}{\mu_0(E)}
\label{MCGR}
\end{equation}
of the number of excited particles, and the corresponding mean-square
fluctuation
\begin{equation}
\langle \delta^2 N_{\rm ex} \rangle_{mc} \; = \; \frac{\mu_2(E)}{\mu_0(E)}
    - \left( \frac{\mu_1(E)}{\mu_0(E)} \right)^2 \; .
\label{MCFL}
\end{equation}
Provided the energy $E$ is so low that a major fraction of the particles
remains in the ground state, i.e., if $\langle N_{\rm ex} \rangle_{mc}$
is sufficiently small as compared to $N$, we can again employ the
oscillator approximation~(\ref{OSAP}), and compute the microcanonical
moments $\mu_k(E)$ from the easily accessible canonical moments~(\ref{CNKM})
by means of saddle-point inversions.

To see in detail how this works, let us first carry through this program
for the paradigmatic example of an isolated ideal Bose gas trapped by a
one-dimensional harmonic oscillator potential. For ease of notation, we
introduce the dimensionless variables $a = \beta\hbar\omega$, characterizing
the inverse temperature, and $n = E/(\hbar\omega)$, corresponding to the total
number of excitation quanta. We are thus working in the regime $a \ll 1$,
$n \gg 1$. Again, this is compatible with the presence of a condensate if
the particle number $N$ is large. Writing $\mu_k(n)$, $Z(a)$, $S_1(a)$,
and $S_2(a)$ instead of $\mu_k(E)$, $Z(\beta)$, $S_1(\beta)$, and $S_2(\beta)$
(see Eqs.~(\ref{CN0M}) -- (\ref{CN2M})), and defining
\begin{eqnarray}
H_0(a) & \equiv & 1
\\
H_1(a) & \equiv & S_1(a)
\label{H1OA}	\\
H_2(a) & \equiv & S_1^2(a) + S_2(a)	\; ,
\label{H2OA}
\end{eqnarray}
the inversion formula acquires the form~\cite{Bethe37}
\begin{equation}
\mu_k(n) \; = \; \left. \frac{e^{na} \, Z(a) \, H_k(a)}
    {\left(-2\pi\frac{\partial n}{\partial a}\right)^{1/2}}
    \right|_{a = a_k(n)}          
\label{BEIV}
\end{equation}
for $k = 0$, $1$, and $2$. It is crucial to note that each moment
requires its own saddle-point parameter $a_k(n)$, obtained by inverting
the corresponding saddle-point equation
\begin{equation}
n \; = \; -\frac{\mbox d}{{\mbox d a}} \ln Z(a)
          -\frac{\mbox d}{{\mbox d a}} \ln H_k(a) \; .
\label{SPE1}
\end{equation}  

In contrast to the canonical case, we now need to evaluate 
the partition sum
\begin{equation}
Z(a) \; = \; \prod_{\nu = 1}^{\infty} \frac{1}{1 - \exp(-a\nu)}
\end{equation}
for $a \ll 1$. This partition sum actually is a well-studied object
in the theory of modular functions; it satisfies a fairly interesting
functional equation that allows one to extract the desired
small-$a$-behaviour straight away~\cite{HardyRamanujan18,GH97c}.
In view of the intended transfer of the method to other trap types,
we refrain from using this particular functional equation here, and resort
once more to the Mellin--Barnes techniques. In this way we get
\begin{eqnarray}
\ln Z(a) & = & -\sum_{\nu = 1}^{\infty} \ln(1 - e^{-a\nu})
    \nonumber	\\
    & = & \frac{1}{2\pi i} \int_{\tau-i\infty}^{\tau+i\infty} \!
       {\mbox d} t \, a^{-t} \Gamma(t) \zeta(t) \zeta(t+1)  \; ,
\label{HRPS}
\end{eqnarray}
so that the dominant pole at $t = 1$ gives the approximation
$\ln Z(a) \approx \zeta(2)/a$. However, in order to derive a proper
asymptotic formula for $Z(a)$ we have to expand $\ln Z(a)$ up to terms
of the order $O(a^0)$ inclusively, which necessitates to take into account
also the double pole of the integrand~(\ref{HRPS}) that lies at $t = 0$.
Since the residue of this pole reads
\begin{equation}
-\zeta(0) \ln a + \zeta'(0) \; = \;
    \frac{1}{2}\ln a - \frac{1}{2}\ln 2\pi  \; , 
\end{equation}
we obtain the desired approximation
\begin{equation}
\ln Z(a) \; = \; \frac{\zeta(2)}{a} + \frac{1}{2}\ln \frac{a}{2\pi}
    + O(a)  \; .
\end{equation}
With $H_1(a)$ and $H_2(a)$ as determined by Eqs.~(\ref{GRD1})
and (\ref{FLD1}), the equations~(\ref{SPE1}) for the saddle-point
parameters then adopt the form  
\begin{equation}
n \; = \; \frac{\zeta(2)}{a^2} + \frac{c(k)}{a}	\; , 
\end{equation}
where
\begin{eqnarray}
c(0) & = & -1/2		\nonumber	\\
c(1) & = & -1/2 + 1	\nonumber	\\
c(2) & = & -1/2 + 2	\; .
\end{eqnarray}
Inverting up to the required order $O(n^{0})$, one finds
\begin{equation}
\frac{1}{a_k(n)} \; = \; \frac{\sqrt{6n}}{\pi}
    - \frac{3 \, c(k)}{\pi^2}   \; ,
\label{TE1D}
\end{equation} 
hence
\begin{equation}
n a_k(n) + \ln Z(a_k(n)) \; = \; \pi\sqrt{\frac{2n}{3}}
    + \frac{1}{2}\ln\!\left(\frac{1}{2\sqrt{6n}}\right)  \; .
\label{HREX}
\end{equation}
The important point to observe here is that the moment-dependent
number $c(k)$ {\em drops out\/}, so that the factors
$e^{n \, a_k(n)}Z(a_k(n))$ entering the inversion formula~(\ref{BEIV})
become asymptotically equal for all $k$. Moreover, also the saddle-point
corrections
\begin{equation}
\left. \left(-2\pi\frac{\partial n}{\partial a}\right)^{-1/2}
    \right|_{a = a_k(n)}
    = \; \sqrt{\frac{3}{2}} \, (6n)^{-3/4}
\label{SPCR}
\end{equation} 
do not develop a significant $k$-dependence, so that these parts cancel
when forming the ratio~(\ref{MCGR}). Hence, we arrive at
\begin{equation}
\frac{\mu_1(n)}{\mu_0(n)} \; = \; H_1(a_1(n)) \; = \; S_1(a_1(n)) \; , 
\end{equation}
indicating that the microcanonical expectation value for the number of
excited particles becomes equal to the canonical expression~(\ref{NEXS})
in the asymptotic regime, where the difference between the saddle-point
parameter $a_1(n)$ and the true inverse temperature $a_0(n)$ is negligible.  
Utilizing the asymptotic temperature--energy relation 
$k_BT/(\hbar\omega) = \sqrt{6n}/\pi$ obtained from Eq.~(\ref{TE1D}),
the microcanonical counterpart to Eq.~(\ref{GRD1}) reads  
\begin{equation}
\langle N_{\rm ex} \rangle_{mc} \; = \; \frac{\sqrt{6n}}{\pi}
    \left[\ln\!\left(\frac{\sqrt{6n}}{\pi}\right) + \gamma \right] \; .
\label{MGR1}
\end{equation}

The calculation of the microcanonical condensate fluctuations requires
more care. The canonical expression~(\ref{FLUS}) had been an immediate
consequence of the definition~(\ref{CNFL}) and Eqs.~(\ref{CN0M}) --
(\ref{CN2M}), relying on the cancellation
$[S_1^2(\beta) + S_2(\beta)] - S_1^2(\beta) = S_2(\beta)$,
but now two different saddle-point parameters enter into the
corresponding difference
\begin{equation}  
H_2(a_2(n)) - H_1^2(a_1(n)) \; = \;
S_2(a_2(n)) + \left(S_1^2(a_2(n)) - S_1^2(a_1(n)) \right) \; ,
\label{DIOH}
\end{equation}
indicating that the microcanonical fluctuation might deviate from the
canonical one. How\-ever, for the one-dimensional oscillator trap we find that
$S_1^2(a_2(n)) - S_1^2(a_1(n)$ is merely of the order $O(\sqrt{n}\ln^2\! n)$,
and thus asymptotically negligible in comparison to $S_2(a_2(n)) = n$.
Therefore, for large $n$ we have
\begin{equation}
\frac{\mu_2(n)}{\mu_0(n)} - \left(\frac{\mu_1(n)}{\mu_0(n)} \right)^2 
    \; = \; S_2(a_2(n))	\; ,
\end{equation}
meaning
\begin{equation}
\langle \delta^2 N_{\rm ex} \rangle_{mc} \; = \; n \; ,
\label{MFL1}
\end{equation}
so that the microcanonical fluctuation of the number of excited Bose
particles in a one-dimensional harmonic trap, considered for energies
such that on the average a fraction of the particles stays in the ground
state, {\em coincides\/} asymptotically with the canonical
fluctuation~(\ref{FLD1}).
     
This analysis can directly be translated into the language of the
theory of partitions of integer numbers~\cite{Andrews76}. Distributing
$n$ excitation quanta among ideal Bose particles, stored in a one-dimensional
harmonic trap, is tantamount to partitioning the number $n$ into summands;
a particle occupying the $\nu$-th excited oscillator state gives a summand
of magnitude $\nu$. In fact, utilizing Eqs.~(\ref{HREX}) and~(\ref{SPCR})
for computing $\mu_0(n)$ according to Eq.~(\ref{BEIV}), one finds
\begin{equation}
\mu_0(n) \; = \; \frac{1}{4\sqrt{3}n}
    \exp\!\left(\pi\sqrt{\frac{2n}{3}}\right)	\; ,
\label{HRPN}
\end{equation}
which is just the celebrated Hardy--Ramanujan formula for the total
number of {\em unrestricted\/} partitions of $n$~\cite{HardyRamanujan18},
corresponding to the number of microstates that are accessible to the Bose
particles when their common excitation energy is $n\hbar\omega$, provided
that $n$ does not exceed the particle number~$N$. For higher energies,
the number of microstates equals the number of partitions that are
{\em restricted\/} by the requirement that there be no more than $N$ summands,
since the $n$~quanta cannot be distributed over more than the $N$
available particles. However, as long as the distribution~(\ref{MCDI})
remains sharply peaked around some value
$\widehat{N}_{\rm ex} \approx \langle N_{\rm ex} \rangle_{mc } <  N$,
even though $n$ may be substantially larger than $N$, the difference between
the number of these restricted and that of the unrestricted partitions is
insignificant, and neglecting this difference is nothing but the oscillator
approximation~(\ref{OSAP}) for $k = 0$.

From the viewpoint of partition theory, Eq.~(\ref{MGR1}) gives
the expected number of summands in a partition of $n$ --- randomly
partitioning $n = 1000$, for instance, one expects roughly $93$ summands
--- and Eq.~(\ref{MFL1}) contains the remarkable statement that the
r.m.s.-fluctuation of the number of parts into which $n$ can be decomposed
becomes just $\sqrt{n}$ in the asymptotic limit~\cite{GH97b}. Higher
cumulants $\kappa_m(n)$ of the distribution which describes the number of
summands in unrestricted partitions of $n$ can be obtained by following the
same strategy as outlined above for calculating $\kappa_2(n) = n$,
leading to, e.g.,   
\begin{equation}
\kappa_3(n) \; = \; \frac{12\sqrt{6} \, \zeta(3)}{\pi^3} \, n^{3/2}
\end{equation} 
and
\begin{equation}
\kappa_4(n) \; = \; \frac{12}{5} \, n^{2}   \; .
\end{equation}
This indicates deviations from a Gaussian distribution, for which
all cumulants higher than the second are zero. A general asymptotic formula
for the partition moments $\mu_k(n)$ for arbitrary~$k$, together with a check
of this formula against exact data for $k = 0, \ldots, 3$, can be found in
Appendix~\ref{AC}.

It is conceptually important to know that the fluctuation formula~(\ref{MFL1})
for the harmonically trapped one-dimensional Bose gas can also be obtained
{\em without\/} invoking the oscillator approximation, since the number of
restricted partitions, and hence the entire distribution~(\ref{MCDI}), can be
well approximated with the help of an asymptotic expression due to Erd\"os and
Lehner~\cite{ErdosLehner41,GH96}. However, that approach depends on a specific
asymptotic result that applies to the one-dimensional harmonic trap only,
whereas the present method is capable of some generalization.

The preceding microcanonical analysis of the ideal Bose--Einstein condensate
in a one-dimensional oscillator trap might appear like much ado about nothing:
we have been careful to keep track of three slightly different saddle-point
parameters, but in the end this distinction turned out to be insignificant,
and we have merely recovered the canonical results. But this is not true in
general; the following reasoning will show that (and why) a condensate in a
three-dimensional isotropic harmonic trapping potential behaves differently.
In this case we again face a partition-type problem, since the total
excitation energy $E$ remains an integer multiple of a basic quantum
$\hbar\omega$. We can then virtually retrace the steps that have led for
$d = 1$ to the microcanonical formulas~(\ref{MGR1}) and (\ref{MFL1}):
Starting from the partition sum
\begin{equation}
Z(a) \; = \; \prod_{\nu = 1}^{\infty}
    \frac{1}{[1 - \exp(-a\nu)]^{(\nu+1)(\nu+2)/2}}
\end{equation}    
and applying the Mellin--Barnes transformation, one readily
finds~\footnote{This expansion had already been derived in 1951 by
    V.S. Nanda with the help of the Euler--Maclaurin summation formula,
    see Ref.~\cite{Nanda51}. The Mellin--Barnes approach followed in the
    present work is {\em much\/} simpler, since it provides immediate access
    to the analytically continued Riemann Zeta function.}
\begin{eqnarray}
\ln Z(a) & = & \frac{\zeta(4)}{a^3} + \frac{3}{2}\frac{\zeta(3)}{a^2}
    + \frac{\zeta(2)}{a} + \frac{5}{8}\ln a
    \nonumber   \\  & &
    - \frac{1}{2}\ln 2\pi + \frac{3}{2}\zeta'(-1)
    + \frac{1}{2}\zeta'(-2) + O(a)  \; ,
\end{eqnarray}
which, together with the canonical expressions~(\ref{GRD3}) and (\ref{FLD3})
that now define $H_1(a)$ and $H_2(a)$ via Eqs.~(\ref{H1OA}) and (\ref{H2OA}),
yields the saddle-point equations
\begin{equation}
n \; = \; \frac{3\zeta(4)}{a^4} + \frac{3\zeta(3)}{a^3}
    + \frac{\zeta(2)}{a^2} + \frac{c(k)}{a} 
\label{SPE3}
\end{equation}
with
\begin{eqnarray}
c(0) & = & -5/8         \nonumber       \\
c(1) & = & -5/8 + 3     \nonumber       \\
c(2) & = & -5/8 + 6     \; .
\end{eqnarray}
Again, these equations differ only to the order $O(a^{-1})$, and
again the moment-dependent coefficient $c(k)$ drops out when computing
$e^{n \, a_k(n)}Z(a_k(n))$ up to the asymptotically relevant terms of
order $O(n^0)$:
\begin{eqnarray}
n a_k(n) + \ln Z(a_k(n)) & = &
    4\zeta(4) \left(\frac{n}{3\zeta(4)}\right)^{3/4}
  + \, \frac{3}{2}\zeta(3) \left(\frac{n}{3\zeta(4)}\right)^{1/2}
    \nonumber	\\  & & 
  + \left[\zeta(2) - \frac{3}{8} \frac{\zeta(3)^2}{\zeta(4)}\right] \!
              \left(\frac{n}{3\zeta(4)}\right)^{1/4}
  - \, \frac{5}{32} \ln\!\left(\frac{n}{3\zeta(4)}\right)
    \nonumber	\\  & &	      
  + \, \frac{\zeta(3)^3}{8\,\zeta(4)^2}
  - \frac{\zeta(2)\zeta(3)}{4\,\zeta(4)}
  - \frac{1}{2} \ln 2\pi
  + \frac{3}{2}\zeta'(-1) + \frac{1}{2}\zeta'(-2) \; .
\label{CANC}
\end{eqnarray} 
It is clear that this cancellation is quite general. Namely, for a
$d$-dimensional isotropic harmonic trapping potential one has
\begin{equation}
\ln Z(a) \; = \; \frac{\zeta(d+1)}{a^d} + \ldots    \; , 
\label{GENZ}
\end{equation}
and the saddle-point equations become
\begin{equation}
n \; = \; \frac{d\,\zeta(d+1)}{a^{d+1}} + \; \ldots \;
    + \frac{c(k)}{a} \; , 
\label{GENN}
\end{equation}      
hence
\begin{equation}
\frac{1}{a_k(n)} \; = \; \left(\frac{n}{d\,\zeta(d+1)}\right)^{1/(d+1)}
    + \; \ldots \; - \frac{c(k)}{d(d+1)\,\zeta(d+1)} \!
    \left(\frac{n}{d\,\zeta(d+1)}\right)^{-(d-1)/(d+1)}	\; . 
\end{equation}
When computing $n a_k(n) + \ln Z(a_k(n))$, the product $n a_k(n)$
contributes a term $c(k)$ that originates from the $O(a^{-1})$-term in
Eq.~(\ref{GENN}). Otherwise, relevant $k$-dependent contributions enter into
$n a_k(n) + \ln Z(a_k(n))$ only via the leading terms of order $O(a^{-d})$
that stem from Eq.~(\ref{GENZ}) on the one hand, and Eq.~(\ref{GENN})
multiplied by $a$ on the other, summing up to
\begin{eqnarray}
(d+1)\,\zeta(d+1) & \times &
    \left[d \left(\frac{n}{d\,\zeta(d+1)}\right)^{(d-1)/(d+1)} \right]
    \nonumber \\
    & \times & 
    \left[ \frac{-c(k)}{d(d+1)\,\zeta(d+1)}
    \left(\frac{n}{d\,\zeta(d+1)}\right)^{-(d-1)/(d+1)} \right]
    \; = \; -c(k) 
\end{eqnarray}
and thus annihilating $c(k)$.

This little calculation, together with the inversion formula~(\ref{BEIV}),
shows explicitly that the microcanonical expectation values
$\langle N_{\rm ex} \rangle_{mc}$ for $d$-dimensional isotropic harmonic
traps become asymptotically equal to their canonical counterparts:
Forming the ratio~(\ref{MCGR}), the factors
$e^{na} Z(a) (-2\pi \partial n/\partial a)^{-1/2}$
cancel even when evaluated at the slightly different saddle-point
parameters $a_1(n)$ and $a_0(n)$, so that we are left with
\begin{equation}
\langle N_{\rm ex} \rangle_{mc} \; = \; S_1(a_1(n)) \; .     
\end{equation}
The asymptotic equality of $\langle N_{\rm ex} \rangle_{mc}$ and
$\langle N_{\rm ex} \rangle_{cn}$ then follows by observing that $a_1(n)$
becomes asymptotically equal to the true inverse temperature $a_0(n)$.  

The computation of the microcanonical condensate fluctuation along
these lines, however, is a much more delicate matter. Returning to the
particular example $d = 3$ for the sake of definiteness, both canonical
expectation values $\langle N_{\rm ex} \rangle_{cn} = S_1(a)$ and
$\langle \delta^2 N_{\rm ex} \rangle_{cn} = S_2(a)$ are determined
by the {\em same\/} simple pole of $Z(\beta,t)$ at $t = 3$, which means
that both $S_1(a)$ and $S_2(a)$ are proportional to $a^{-3}$. This, in turn,
implies that in contrast to the one-dimensional case the difference 
$S_1^2(a_2(n)) - S_1^2(a_1(n))$ appearing in Eq.~(\ref{DIOH}) now is
of the {\em same\/} order as $S_2(a_2(n))$ itself, so that here the
quite innocent-looking difference between the saddle-point
equations~(\ref{SPE3}), even though only of the order $O(a^{-1})$ and
apparently hidden behind terms of order $O(a^{-4})$ --- thus being
overwhelmed much stronger than for $d = 1$ --- {\em must\/} lead to an
asymptotic difference between canonical and microcanonical condensate
fluctuations: The exponent of $T$ will be the same, but the prefactors
will differ. This observation forces us to evaluate Eq.~(\ref{MCFL})
very carefully. We may not simply rely on the cancellation of $c(k)$
as found in Eq.~(\ref{CANC}), but have to expand both ratios
$\mu_2(n)/\mu_0(n)$ and $\mu_1(n)/\mu_0(n)$ consistently up to terms
of order $O(a_k(n)^{-3}) = O(n^{3/4})$. This forces us to expand
$n a_k(n) + \ln Z(a_k(n))$, as well as the saddle-point corrections, even
up to terms of the order $O(n^{-3/4})$! Detailed analysis shows that such an
expansion is possible even with only the saddle-point equations~(\ref{SPE3})
as input, although this is not immediately obvious. Proceeding in this manner,
we find 
\begin{eqnarray}
\langle \delta^2 N_{\rm ex} \rangle_{mc} & = &
    \left[1 + \frac{33}{12\,\zeta(4)}
	\left(\frac{3\zeta(4)}{n}\right)^{3/4} \right] H_2(a_2(n))
  - \left[1 + \frac{2}{\zeta(4)}
        \left(\frac{3\zeta(4)}{n}\right)^{3/4} \right] H_1^2(a_1(n))
  \nonumber \\
& = & S_2(a_2(n)) + \left(S_1^2(a_2(n)) - S_1^2(a_1(n))\right)
  + \frac{3}{4}\frac{\zeta(3)^2}{\zeta(4)}
  \left(\frac{n}{3\zeta(4)}\right)^{3/4} \; .   
\label{MSP3}
\end{eqnarray}
The $O(n^{-3/4})$-corrections in the square brackets arise because the
inverse temperature $a_0(n)$ differs from the saddle-point parameters
$a_2(n)$ and $a_1(n)$; this is what causes the last term in the second
equation. Since, moreover,
\begin{equation}
S_1^2(a_2(n)) - S_1^2(a_1(n)) \; = \;
  - \frac{3}{2}\frac{\zeta(3)^2}{\zeta(4)}
  \left(\frac{n}{3\zeta(4)}\right)^{3/4}    \; ,
\end{equation}
we finally arrive at
\begin{eqnarray}
\langle \delta^2 N_{\rm ex} \rangle_{mc} & = &
    \left[ \zeta(2) - \frac{3}{4}\frac{\zeta(3)^2}{\zeta(4)} \right]
    \left(\frac{n}{3\zeta(4)}\right)^{3/4}
    \nonumber	\\
    & = &
    \left[ \zeta(2) - \frac{3}{4}\frac{\zeta(3)^2}{\zeta(4)} \right]
    \left(\frac{k_BT}{\hbar\omega}\right)^3 \; .
\label{MFL3}
\end{eqnarray}
This quantifies what we have anticipated: Apparently tiny differences between
the three saddle-point parameters conspire to lower the microcanonical
mean-square condensate fluctuation against the canonical result~(\ref{FLD3}),
as a consequence of the fact that the rightmost pole of $Z(\beta,t)$
governs both $\langle N_{\rm ex} \rangle_{cn} = S_1(\beta)$ and
$\langle \delta^2 N_{\rm ex} \rangle_{cn} = S_2(\beta)$. For the
one-dimensional harmonic trap, where $\langle N_{\rm ex} \rangle_{cn}$ and
$\langle \delta^2 N_{\rm ex} \rangle_{cn}$ are determined by two different
poles, such an asymptotic difference does not exist.

Conceptually instructive as the above calculation may be, it is also lacking
elegance, to say the least. The reason for the appearance of cumbersome
equations like Eq.~(\ref{CANC}) or Eq.~(\ref{MSP3}) lies in the fact that one
exctracts the fluctuations from the exponentially large moments $\mu_k(E)$,
taking the difference~(\ref{MCFL}). This involves huge cancellations, as
becomes dramatically clear already for $d = 1$ by comparing the numbers listed
in Tables~I and~V of Appendix~\ref{AC}. If one could avoid computing the
microcanonical moments, and aim directly for the {\em difference\/} between
canonical and microcanonical fluctuations, one should get expressions of a
far simpler nature. The following section will show that such a strategy
is actually feasible.

\section{Mellin-Barnes approach to microcanonical condensate fluctuations}
\label{S5}

We start by considering the excited-states subsystem with the fugacity~$z$
and the energy $E$ as basic variables~\cite{NavezEtAl97}, so that we have
the relation $N_{\rm ex} = N_{\rm ex}(z,E)$. In principle, $N_{\rm ex}$
depends also on trap parameters that determine the single-particle energies,
like the oscillator frequencies in the case of harmonic traps, but these
parameters will be kept constant in the following. Taking the total
differential, 
\begin{equation}
{\mbox d} N_{\rm ex} \; = \;
    \left(\frac{\partial N_{\rm ex}}{\partial z}\right)_{\!\! E}{\mbox d} z +
    \left(\frac{\partial N_{\rm ex}}{\partial E}\right)_{\!\! z}{\mbox d} E \; ,
\end{equation}
then keeping the temperature $T$ fixed, one finds
\begin{equation}
\left. z \left(\frac{\partial N_{\rm ex}}{\partial z}\right)_{\!\! T}
    \, \right|_{z=1} \; = \;
    z \left[ \left(\frac{\partial N_{\rm ex}}{\partial z}\right)_{\!\! E}
    + \left(\frac{\partial N_{\rm ex}}{\partial E}\right)_{\!\! z}
      \left(\frac{\partial E}{\partial z}\right)_{\!\! T} \right]_{z=1} \; .  
\end{equation}
The left hand side equals the canonical mean-square fluctuation
$\langle \delta^2 N_{\rm ex} \rangle_{cn}$, whereas the first term on the
r.h.s.\ is its microcanonical counterpart
$\langle \delta^2 N_{\rm ex} \rangle_{mc}$. Hence, we obtain
\begin{eqnarray}
\langle \delta^2 N_{\rm ex} \rangle_{cn} -
    \langle \delta^2 N_{\rm ex}\rangle_{mc} & = & \left.
    \left(\frac{\partial N_{\rm ex}}{\partial E}\right)_{\!\! z}
    \left(\frac{\partial E}{\partial z}\right)_{\!\! T}
    \, \right|_{z=1}	\nonumber   \\      & = &
    \frac{k_BT^2 \left.
    \left(\frac{\partial N_{\rm ex}}{\partial T}\right)_{\!\! z}
    \left(\frac{\partial E}{\partial z}\right)_{\!\! T}
    \, \right|_{z=1} }
    {k_BT^2 \left.
    \left(\frac{\partial E}{\partial T}\right)_{\!\! z}
    \, \right|_{z=1} }	\; .
\end{eqnarray}
Now the denominator 
\begin{equation}
k_BT^2 \left. \left(\frac{\partial E}{\partial T}\right)_{\!\! z}
    \, \right|_{z=1} \; = \; \langle \delta^2 E \rangle_{cn} 
\end{equation}
is the canonical mean-square fluctuation of the system's energy,
whereas the two partial derivatives in the numerator,
\begin{equation}
k_BT^2 \left. \left(\frac{\partial N_{\rm ex}}{\partial T}\right)_{\!\! z}
    \, \right|_{z=1} \; = \;
    \left. \left(\frac{\partial E}{\partial z}\right)_{\!\! T}
    \, \right|_{z=1} \; = \;
    \langle \delta N_{\rm ex} \, \delta E \rangle_{cn}	\; ,
\end{equation}
both equal the canonical particle-energy correlation
$\langle \delta N_{\rm ex} \, \delta E \rangle_{cn}
= \langle N_{\rm ex} E \rangle_{cn}
- \langle N_{\rm ex} \rangle_{cn} \langle E \rangle_{cn}$.
Thus, we arrive at the noteworthy identity 
\begin{equation}
   \langle \delta^2 n_0 \rangle_{cn} - \langle \delta^2 n_0 \rangle_{mc}
   \; = \;  
   \frac{\left[\langle \delta N_{\rm ex} \, \delta E \rangle_{cn} \right]^2}
   {\langle \delta^2 E \rangle_{cn}} 
\end{equation}
which expresses the difference between canonical and microcanonical
condensate fluctuations in terms of quantities that can be computed
entirely within the convenient canonical ensemble. The usefulness of this
formula, first stated by Navez et al.~\cite{NavezEtAl97}, rests in the fact
that it lends itself again to the oscillator approximation, and thus to an
efficient evaluation by means of the Mellin--Barnes transformation: Within
the oscillator approximation, the canonical particle-energy correlation
becomes 
\begin{eqnarray}
\langle \delta N_{\rm ex} \, \delta E \rangle_{cn}
    & = & \left. \left(z\frac{\partial}{\partial z}\right)
                 \left(-\frac{\partial}{\partial\beta}\right)
          \ln \, \Xi_{\rm ex}(z,\beta) \right|_{z=1}
	  \nonumber   \\
    & = & \sum_{\nu=1}^{\infty}
	  \frac{\varepsilon_{\nu}}{\exp(\beta\varepsilon_\nu) - 1}
          \left( \frac{1}{\exp(\beta\varepsilon_\nu) - 1} + 1 \right)
	  \nonumber   \\
    & = & \frac{1}{\beta} \,
	  \frac{1}{2\pi i}\int_{\tau-i\infty}^{\tau+i\infty} \! {\mbox d} t \,
	  \Gamma(t) Z(\beta,t-1) \zeta(t-1)   \; ,
\end{eqnarray}
and the canonical energy fluctuation adopts the quite similar form
\begin{eqnarray}
\langle \delta^2 E \rangle_{cn}
    & = & \left. \left(-\frac{\partial}{\partial\beta}\right)^{\!2}
          \ln \, \Xi_{\rm ex}(z,\beta) \right|_{z=1}
	  \nonumber   \\
    & = & \sum_{\nu=1}^{\infty}
          \frac{\varepsilon_{\nu}^2}{\exp(\beta\varepsilon_\nu) - 1}
	  \left( \frac{1}{\exp(\beta\varepsilon_\nu) - 1} + 1 \right)
	  \nonumber   \\
    & = & \frac{1}{\beta^2} \,
          \frac{1}{2\pi i}\int_{\tau-i\infty}^{\tau+i\infty} \! {\mbox d} t \,
	  \Gamma(t) Z(\beta,t-2) \zeta(t-1)   \; .
\end{eqnarray}
Hence,    
\begin{equation}
\langle \delta^2 n_0 \rangle_{cn} - \langle \delta^2 n_0 \rangle_{mc} \; = \;
   \frac{\left[ \frac{1}{2\pi i} \int_{\tau-i\infty}^{\tau+i\infty} \!
   {\mbox d} t \, \Gamma(t) Z(\beta,t-1) \zeta(t-1) \right]^2}
   {\frac{1}{2\pi i} \int_{\tau-i\infty}^{\tau+i\infty} \! {\mbox d} t \,
   \Gamma(t) Z(\beta,t-2) \zeta(t-1)}	    \; .
\label{DOFL}
\end{equation}

Compared to the saddle-point approach in the preceding section, this
formula is remarkably easy to handle. Applied to the one-dimensional
harmonic trap, for instance, it yields immediately
\begin{equation}
\langle \delta^2 n_0 \rangle_{cn} - \langle \delta^2 n_0 \rangle_{mc}
\; = \; \frac{1}{2\zeta(2)} \frac{k_BT}{\hbar\omega}
    \left[ \ln \! \left(\frac{k_B T}{\hbar\omega}\right)
         + \gamma + 1 \right]^2 \; .
\end{equation}
Improving Eq.~(\ref{FLD1}) by taking also the next-to-leading pole
into account,
\begin{equation}
\langle \delta^2 n_0 \rangle_{cn} \; = \; 
    \zeta(2) \left(\frac{k_B T}{\hbar\omega}\right)^2
  - \; \frac{1}{2} \, \frac{k_B T}{\hbar\omega}   \; ,
\end{equation}
and substituting $k_BT/(\hbar\omega) = \sqrt{6n}/\pi  + 3/(2\pi^2)$
as stated by Eq.~(\ref{TE1D}), we arrive at a refined approximation
to the microcanonical ground state fluctuation: 
\begin{equation} 
\langle \delta^2 n_0 \rangle_{mc} \; = \; n - \frac{3\sqrt{6n}}{\pi^3}
    \left[ \ln \! \left(\frac{\sqrt{6n}}{\pi}\right)
    + \gamma + 1 \right]^2 \; .
\label{MFI1}
\end{equation}
As we already know, the relative difference between canonical and
microcanonical mean-square fluctuations vanishes asymptotically for a
condensate in a one-dimensional harmonic trap. But still, this relative
difference is of the order $O(\ln^2\! n/\sqrt{n})$, so that Eq.~(\ref{MFI1})
is substantially more accurate than the previous leading-order
approximation~(\ref{MFL1}).

This fluctuation formula~(\ref{MFI1}) again has an interesting
number-theoretical interpretation. As indicated in the previous section,
the oscillator approximation, when applied to a one-dimensional harmonic
trapping potential, corresponds to neglecting the difference between
partitions of $n$ into no more than $N$ summands and unrestricted partitions;
this remains exact as long as $n \le N$. Hence, Eq.~(\ref{MFI1}) provides
{\em a forteriori\/} a fair approximation to the fluctuation of the number
of integer summands into which the integer~$n$ can be decomposed.
Figure~1 depicts the r.m.s-fluctuation
$\sigma(n) = \langle \delta^2 n_0 \rangle_{mc}^{1/2}$ as approximated by
Eq.~(\ref{MFL1}) (upper dashed line) and by Eq.~(\ref{MFI1})
(lower dashed line; coinciding almost with the full line), and compares
these approximations to the exact data (full line). The latter have
been computed numerically from the distribution~(\ref{MCDI}) for the
one-dimensional harmonic trap, utilizing the recursion relation
\begin{equation}
\Phi(N_{\rm ex}|n\hbar\omega) \; = \;
    \sum_{k=1}^{\min(n-N_{\rm ex},N_{\rm ex})}
    \Phi(k\,|(n-N_{\rm ex})\hbar\omega)
\end{equation}
with $\Phi(1\,|1\,\hbar\omega) = \Phi(n|n\hbar\omega) = 1$, assuming
$n \le N$. The agreement between the exact fluctuation and the improved
asymptotic formula is no less than striking. It should be noted that,
within the oscillator approximation, Eq.~(\ref{MFI1}) describes the
fluctuation of the ground state particles not only up to the ``restriction
temperature'' $T_R = (\hbar\omega/k_B)\sqrt{N/\zeta(2)}$, where $n = N$,
but almost up to $T_0 = (\hbar\omega/k_B)N/\ln N$, where the occupation
of the ground state becomes significant~\cite{GH96}.

The handiness of the fluctuation formula~(\ref{DOFL}) becomes fully clear
when dealing with $d$-dimensional harmonic traps, $d \ge 2$. Since isotropic
harmonic traps can be considered as special cases, we proceed at once to 
anisotropic potentials, and consider $d = 2$ first. All the required
technicalities have already been collected in Section~\ref{S3} and
Appendix~\ref{AA}: The integrand in the denominator of Eq.~(\ref{DOFL})
has its rightmost pole at $t = 3$ with residue  
$\Gamma (3) (k_B T/\hbar \Omega )^2 \zeta (2)$; the rightmost pole
of the integrand in the numerator lies at $t = 4$, with residue 
$\Gamma (4) (k_B T/\hbar \Omega )^2 \zeta (3)$. Thus,
\begin{equation}
\langle \delta^2 n_0 \rangle_{cn} - \langle \delta^2 n_0 \rangle_{mc}
    \; = \; \frac{2}{3} \frac{\zeta{(2)}^2}{\zeta(3)}
    \left(\frac{k_BT}{\hbar\Omega}\right)^2 \, ,
\end{equation}
implying that the difference between canonical and microcanonical fluctuations
is still small, even if only by a logarithm, compared to the canonical
fluctuations~(\ref{FLD2}) or~(\ref{delani2}), respectively. Hence, in the
asymptotic limit $k_BT/(\hbar\Omega) \gg 1$ --- assuming $N$ is that large
that this limit can be reached with a condensate --- canonical and
microcanonical condensate fluctuations still agree. But this is clearly
the marginal case, as witnessed by the fact that for $d = 2$ the poles of
$Z(\beta,t)$ and $\zeta(t-1)$ in Eq.~(\ref{MBFL}) fall together. 

For $d \ge 3$, the relevant poles in the integrands of Eq.~(\ref{DOFL}) 
are located at $t = d+1$ and $t = d+2$, and one finds the general formula 
\begin{equation}
\langle \delta^2 n_0 \rangle_{cn} - \langle \delta^2 n_0 \rangle_{mc}
    \; = \; \frac{d}{d+1} \frac{\zeta{(d)}^2}{\zeta(d+1)}
    \left(\frac{k_BT}{\hbar\Omega}\right)^d \, .
\end{equation}
In particular, for isotropic three-dimensional traps one recovers, but now
without any substantial effort, the previous result~(\ref{MFL3}). More
generally, for $d \ge 3$ the condensate fluctuations in harmonically trapped,
energetically isolated ideal Bose gases are significantly smaller than the
corresponding fluctuations~(\ref{delanid}) in traps that are thermally
coupled to some heat bath, although the exponent of $T$ remains the same.
As we have repeatedly emphasized, this finding is explained by the fact
that for $d \ge 3$ both $\langle N_{\rm ex} \rangle_{cn}$ and
$\langle \delta^2 N_{\rm ex} \rangle_{cn}$ are determined by the
same simple pole of the spectral Zeta function $Z(\beta,t)$.

\section{Conclusions}
\label{S6}

The three central formulas explained in this work, Eqs.~(\ref{MBGR}),
(\ref{MBFL}), and (\ref{DOFL}), vindicate the assertion put forward in the
Introduction: It is the location of the rightmost pole of the spectral Zeta
function $Z(\beta,t)$ that determines the statistical properties of the
condensate. If that pole is located at $t = p$, and $0 < p < 1$, then we
deduce from Eq.~(\ref{MBGR}) that $\langle N_{\rm ex} \rangle_{cn}$ grows
linearly with temperature (so that $\langle n_{0} \rangle_{cn}$ decreases
linearly with $T$) in the condensate regime, irrespective of the detailed
properties of the trap, since in that case the pole of $\zeta(t)$ lies to
the right of $p$. If $p = 1$, the poles of $Z(\beta,t)$ and $\zeta(t)$ fall
together, so that the linear temperature dependence develops a logarithmic
correction. If $p > 1$, we have $\langle N_{\rm ex} \rangle_{cn} \propto T^p$.

The key point to be noted when discussing canonical condensate fluctuations
is that the pole of the Riemann Zeta function $\zeta(t-1)$ in Eq.~(\ref{MBFL})
lies at $t = 2$, so that $\langle \delta^2 n_0 \rangle_{cn}$ changes its
$T$-dependence at $p = 2$: If $0 < p < 2$, then
$\langle \delta^2 n_0 \rangle_{cn} \propto T^2$; if $p = 2$, there is the
familiar logarithmic correction to this quadratic $T$-dependence, as expressed
by Eqs.~(\ref{FLD2}) and (\ref{delani2}) for two-dimensional harmonic traps;
if $p > 2$, then $\langle \delta^2 n_0 \rangle_{cn} \propto T^p$.

The saddle-point calculations in Section~\ref{S4} may be cumbersome,
but they exemplify on an elementary level why $\langle n_0 \rangle_{cn}$
equals $\langle n_0 \rangle_{mc}$ in the asymptotic regime, and why canonical
and microcanonical condensate fluctuations may differ. Equation~(\ref{MSP3})
summarizes the essentials for the three-dimensional oscillator trap: One
needs three slightly different saddle-point parameters for computing the
required microcanonical moments~(\ref{MIMO}) within the oscillator
approximation~(\ref{OSAP}) from their canonical counterparts~(\ref{CNKM});
these slight differences lower the microcanonical fluctuation against the
canonical one. The elegant expression~(\ref{DOFL}) links the difference 
$\langle \delta^2 n_0 \rangle_{cn} - \langle \delta^2 n_0 \rangle_{mc}$  
again to the dominant pole of $Z(\beta,t)$: If $0 < p < 2$, that difference
has an exponent of $T$ which is smaller than that of
$\langle \delta^2 n_0 \rangle_{cn}$, so that both types of fluctuations
become asymptotically equal, but if $p > 2$, then the difference acquires
the same exponent of $T$ as $\langle \delta^2 n_0 \rangle_{cn}$, so that
the microcanonical condensate fluctuation remains lower than the canonical
one even in the asymptotic regime.

We have evaluated canonical and microcanonical condensate fluctuations
explicitly for harmonic trapping potentials, where $Z(\beta,t)$ reduces to
the familiar Riemann or Barnes-type Zeta functions. This may appear a bit
special, but an analogous discussion is possible for quite arbitrary traps,
if one merely exploits the connection between the residues of $Z(\beta,t)$
and the corresponding heat-kernel coefficients (see Ref.~\cite{KirstenToms98}
for a brief explanation of this fairly deep connection).
 
The vision of letting the poles of $Z(\beta,t)$ move in the complex
$t$-plane is not a fantasy restricted to the theorist's ivory tower,
but may have direct experimental consequences. Continuously deforming
the trapping potential means continuously changing the trap's
single-particle spectrum, and hence shifting $p$. For example,
a spectrum of the type~\cite{deGrootEtAl50,WeissWilkens97}
\begin{equation}
\varepsilon_{\nu_1,\ldots,\nu_d} \; = \;
    \varepsilon_0 \sum_{i = 1}^d c_i \nu_i^s
\label{DGSP}
\end{equation}
with integer quantum numbers $\nu_i$ and dimensionless anisotropy coefficients
$c_i$ not too different from unity implies $p = d/s$~\cite{GH97b}, so that,
e.g., steepening a three-dimensional harmonic oscillator potential ($s = 1$)
towards a box potential ($s = 2$) means lowering $p$ from $3$ to $3/2$.
During such a process, the fluctuation of a large condensate is described
by an exponent of $T$ that {\em changes\/} as long as $p$ remains above~$2$,
since then $\langle \delta^2 n_0 \rangle_{mc} \propto T^{d/s}$,
but {\em remains constant\/} when $p$ decreases further;
$\langle \delta^2 n_0 \rangle_{mc} \propto T^{2}$.
  
There is another detail that deserves to be mentioned. From the viewpoint
of partition theory, Eq.~(\ref{MFL1}) states that the r.m.s.-fluctuation of
the number of parts into which a large integer~$n$ can be decomposed is
approximately normal, $\sigma(n) \sim \sqrt{n}$. However, when characterizing
condensate fluctuations, one would not do so in terms of the number of
excitation quanta $n$, but rather in terms of the number of excited particles
$\langle N_{\rm ex} \rangle_{mc}$~\footnote{We choose to characterize the
fluctuation in terms of $\langle N_{\rm ex} \rangle$, rather than
$\langle n_0 \rangle$, since the properties of an ideal Bose--Einstein
condensate, including its fluctuation, are independent of
$\langle n_0 \rangle$. This is just what is exploited in the oscillator
approximation.}.
But then Eqs.~(\ref{MGR1}) and (\ref{MFL1}) yield
\begin{equation}
\langle \delta^2 N_{\rm ex} \rangle^{1/2}_{mc} \; \propto \;
    \langle N_{\rm ex} \rangle_{mc}	\; ,  
\end{equation}
apart from logarithmic corrections,
stating that the {\em normal\/} partition-theoretic fluctuation translates
into {\em supranormal\/} fluctuation of the number of excited Bose particles
in a one-dimensional harmonic trap. More generally, for traps with
single-particle spectra~(\ref{DGSP}) one obtains~\cite{GH97b}
\begin{eqnarray}
\langle \delta^2 N_{\rm ex} \rangle^{1/2} \propto
    \langle N_{\rm ex} \rangle\;\;\;\;
    & \qquad \mbox{for} \qquad & 0 < d/s < 1 \; ,   \nonumber	\\
\langle \delta^2 N_{\rm ex} \rangle^{1/2} \propto
    \langle N_{\rm ex} \rangle^{s/d}
    & \qquad \mbox{for} \qquad & 1 < d/s < 2 \; ,   \nonumber	\\
\langle \delta^2 N_{\rm ex} \rangle^{1/2} \propto
    \langle N_{\rm ex} \rangle^{1/2}
    & \qquad \mbox{for} \qquad & 2 < d/s	\; ;
\end{eqnarray}
both within the canonical and the microcanonical ensemble. Hence, when
increasing $d/s$ from the $1d$-harmonic oscillator value~$1$, the degree of
supranormality is gradually lowered, until one arrives at normal particle
number fluctuations for $d/s > 2$. 
  
Taking these insights together with those obtained in the related previous
works~\cite{GH96,Politzer96,GajdaRzazewski97,WilkensWeiss97,WeissWilkens97,NavezEtAl97,GH97b},
it seems fair to conclude that by now a classic problem in statistical
mechanics, the fluctuation of an ideal Bose--Einstein condensate, has been
fully understood.

\appendix

\section{Residues of Barnes-type Zeta functions}
\label{AA}

According to Section~\ref{S3}, the spectral Zeta function for
$d$-dimensional anisotropic harmonic traps adopts the Barnes form
\begin{equation}
Z(\beta, t) \; = \sum_{\vec \nu \in \nats_0^d / \{0\}}
    \frac{1}{(\beta \hbar \vec \omega \vec \nu)^t }   \; ,
\label{specami}
\end{equation}
and the canonical thermodynamics of an ideal Bose gas stored in such a trap
depends crucially on the rightmost pole of this function. In this appendix
we briefly sketch the derivation of Eq.~(\ref{resbarnes}), i.e.,
of the residue of the rightmost pole of $Z(\beta,t)$.

The starting point is the contour integral representation of the
Gamma function~\cite{GR80},
\begin{equation}
\Gamma(t) \; = \; \frac{i}{2\sin(\pi t)}
    \int_{{\cal C}} \! {\mbox d} \alpha \, (-\alpha)^{t-1} e^{-\alpha} \; ,
\label{A2}
\end{equation}
where ${\cal C}$ is enclosing the positive real axis counterclockwise.
With the help of this representation we deduce
\begin{eqnarray} 
Z(\beta, t) & = & \frac{i}{2\sin(\pi t) \, \Gamma(t)}
    \int_{{\cal C}} \! {\mbox d} \alpha \, (-\alpha)^{t-1} e^{-\alpha}
    \sum_{\vec \nu \in \nats_0^d / \{0\}}
    \frac{1}{(\beta \hbar \vec \omega \vec \nu)^t }   \nonumber \\
	    & = & \frac{i}{2\sin(\pi t) \, \Gamma(t)}
    \sum_{\vec \nu \in \nats_0^d / \{0\}}
    \int_{{\cal C}} \! {\mbox d} \alpha \, (-\alpha)^{t-1}
    e^{-\alpha\beta\hbar\vec\omega\vec \nu}	    \nonumber \\
	    & = & -\frac{\Gamma (1-t)}{2\pi i}
    \int_{{\cal C}} \! {\mbox d} \alpha \, (-\alpha)^{t-1} 
    \left\{\frac{1}{\prod_{i=1}^d
	\left(1 - e^{-\alpha\beta\hbar \omega_i}\right)} -1 \right\} \; .
\label{A3}
\end{eqnarray}
The first equality is obtained by interchanging summation and integration,
then changing in each summand from the integration variable $\alpha$ to
$\alpha\beta\hbar\vec\omega\vec\nu$; the second by summing the resulting
geometric series and utilizing the relation
\begin{equation}
\sin(\pi t) \, \Gamma(t) \; = \; \frac{\pi}{\Gamma(1-t)} \; .
\end{equation}  

The poles of $Z(\beta, t)$ are featured by Eq.~(\ref{A3}) in a particularly
transparent manner. Namely, the prefactor $\Gamma(1-t)$ has simple poles at
integer values $t = 1,2,3,\ldots \,$. At these values the remaining contour
integral may be evaluated immediately by just collecting the residues
enclosed by ${\cal C}$. The only possible pole contributing to the integral
lies at $\alpha = 0$; it has nonvanishing residues for
$t = -\infty, \ldots, -1,0,1, \ldots, d$. Hence, the poles of $Z(\beta,t)$
are located at $t = 1,\ldots,d$, and the residue of the rightmost pole is
found to be
\begin{equation}
\mbox{res} \; Z(\beta, d) \; = \; (-1)^{d-1}
    \prod_{i=1}^d \left(\beta\hbar\omega_i\right)^{-1} \, 
    \mbox{res} \; \Gamma(1-d)  \; .
\end{equation}
Using the identity
\begin{equation} 
\mbox{res} \; \Gamma(-n) \; = \; \frac{(-1)^n} {n!} \; ,
    \qquad n \in \nats_0 \; ,
\end{equation}
we arrive directly at Eq.~(\ref{resbarnes}).

\section{Canonical condensate fluctuation for $d=2$}
\label{AB}

When evaluating the fluctuation formula~(\ref{MBFL}) for two-dimensional
harmonic traps, the product $Z(\beta,t) \, \zeta(t-1)$ provides a double
pole at $t = 2$. In that case the knowledge of the residue~(\ref{resbarnes})
is not enough for computing the mean-square condensate fluctuation;
also the finite part of $Z(\beta,t)$ at $t = 2$ enters into the residue of
the double pole. More precisely, in analogy to Eq.~(\ref{riemann}) for the
Riemann Zeta function, one needs the expansion  
\begin{equation}
Z(\beta,t) \; = \; \left(\frac{k_B T}{\hbar \Omega}\right)^2
    \left(\frac{1}{t-2} + f(\omega_1, \omega_2 , t) \right)
\end{equation}
for $t$ close to $2$. In this appendix we determine the function
$f(\omega_1, \omega_2, t)$, and thus prove Eq.~(\ref{delani2}).

Introducing $a= \sqrt{\omega_1/\omega_2}$ and $b=\sqrt{\omega_2/\omega_1}$,
we first write
\begin{equation}
Z(\beta,t) \; = \; \left(\frac{k_B T}{\hbar \Omega}\right)^t
    \sum_{\vec \nu \in \nats_0^2 / \{0\}} \frac{1}{(a\nu_1+b\nu_2)^t} \; ,
\end{equation}
valid for $R\!e\,(t) > 2$. Splitting the sum according to the scheme
\[
\sum_{\vec \nu \in \nats_0^2 / \{0\}} \; = \;
    \sum_{\nu_1 = 1}^{\infty} (\nu_2 = 0)
  + \sum_{\nu_2 =1} ^{\infty} (\nu_1 =0 ) \; 
  + \sum_{\nu_1,\nu_2 = 1}^{\infty} \; ,
\]
we find the decomposition
\begin{equation}
Z(\beta,t) \; = \; \left(\frac{k_B T}{\hbar \Omega}\right)^t
    \left\{ \zeta(t) (a^{-t} + b^{-t})
  + H(\omega_1, \omega_2, t) \right\}	\; ,
\label{B3}
\end{equation}
where
\begin{eqnarray}
H(\omega_1, \omega_2, t) & = & \sum_{\nu_1,\nu_2 = 1}^{\infty}
    \frac{1}{(a\nu_1+b\nu_2)^t}	\nonumber   \\
    & = & \frac{1}{\Gamma(t)}
    \int_0^{\infty} \! {\mbox d} \alpha \, \alpha ^{t-1}
    \frac{e^{-(a+b)\alpha}}
         {\left(1-e^{-a\alpha}\right)\left(1-e^{-b\alpha}\right)} \; .
\label{B4}
\end{eqnarray} 
This identity is obtained in a similar manner as Eq.~(\ref{A3}), using
the familiar representation 
\begin{equation}
\Gamma(t) \; = \; \int_0^{\infty} \! {\mbox d}\alpha \,
    \alpha^{t-1} e^{-\alpha}
\label{B5}
\end{equation}
of the Gamma function.
    
Now we are interested in the behaviour of $Z(\beta,t)$ as $t\to 2$, where,
as we know from Appendix~\ref{AA}, it has a simple pole. How is this
realized in Eq.~(\ref{B3})? Since $\zeta(t)$ is regular at $t=2$, the
pole is contained in the integral~(\ref{B4}). At the lower integration
bound, that is, for $\alpha \to 0$, the integrand behaves as $1/\alpha$
for $t\to 2$; therefore the integral diverges at $t = 2$. The behaviour
of the integral as $t$ tends to $2$ is extracted with the help of the
following trick. For $R\!e\,(t) > 2$, write 
\begin{eqnarray}
H(\omega_1, \omega_2, t) & = & \frac{1}{\Gamma(t)}
     \int_0^{\infty} \! {\mbox d}\alpha \, \alpha^{t-1} e^{-(a+b)\alpha}
     \left(\frac{1}{\left(1-e^{-a\alpha}\right) 
                    \left(1-e^{-b\alpha}\right)}
   - \frac{1}{\alpha^2} + \frac{1}{\alpha^2} \right)	\nonumber  \\
   & = &   \frac{\Gamma(t-2)}{\Gamma(t)}(a+b)^{2-t}	\nonumber  \\
   &   & + \; \frac{1}{\Gamma(t)}
     \int_0^{\infty} \! {\mbox d}\alpha \, \alpha^{t-1} e^{-(a+b)\alpha}
     \left(\frac{1}{\left(1-e^{-a\alpha}\right)
		    \left(1-e^{-b\alpha}\right)}
   - \frac{1}{\alpha^2}	\right)	    \; ,
              \nonumber 
\end{eqnarray}
where Eq.~(\ref{B5}) has been used. The simple pole of $Z(\beta,t)$ 
at $t=2$ is now contained in the first term, since 
$\Gamma(t-2) / \Gamma(t) = 1/[(t-1)(t-2)]$, and the remaining integral
is finite for $t = 2$. In this way, we arrive at the expansion
\begin{equation}
H(\omega_1, \omega_2, t) \; = \; \frac{1}{t-2} - 1 -
    \ln\!\left(\sqrt{\frac{\omega_1}{\omega_2}} + 
               \sqrt{\frac{\omega_2}{\omega_1}}\right)
  + I(\omega_1,\omega_2) + O(t-2)   \; ,
\end{equation}
with $I(\omega_1,\omega_2)$ as defined in Eq.~(\ref{finiteb}). Together
with Eq.~(\ref{B3}), this determines the desired function
$f(\omega_1,\omega_2,t)$ and thereby leads to the result~(\ref{delani2}).

It is quite interesting to see how the fluctuation formula~(\ref{FLD2})
for the isotropic case is recovered in the limit
$\omega_1 = \omega_2 = \omega$. Then the integral simplifies to 
\begin{eqnarray}
I(\omega,\omega ) & = & \int_0^{\infty} \! {\mbox d} \alpha \, 
    \alpha e^{-2\alpha} \left(\frac{1}{\left(1-e^{-\alpha}\right)^2} 
  - \frac{1}{\alpha^2}\right)	    \nonumber	\\
  & = & 2 - \int_0^{\infty} \! {\mbox d} \alpha \, \left[
    e^{-2\alpha}\left(\frac{1}{\alpha} - \frac{1}{1 - e^{-\alpha}} \right)
    + \frac{\alpha e^{-\alpha}}{1 - e^{-\alpha}} \right] \; .
\end{eqnarray}
Employing now the identities~\cite{GR80}
\begin{equation}
\psi(z) \; = \; \frac{\mbox{d}}{\mbox{d}z} \ln \Gamma(z)
	\; = \; \ln z + \int_0^{\infty} \! {\mbox d}\alpha \, e^{-z\alpha}
	\left( \frac{1}{\alpha} - \frac{1}{1 - e^{-\alpha}}\right)
\end{equation}
for the Psi function, and
\begin{equation}
\zeta_H(z,q) \; = \; \sum_{n=0}^{\infty} \frac{1}{(n+q)^z} 
        \; = \; \frac{1}{\Gamma (z)} \int_0^{\infty} \! {\mbox d}\alpha \,
	 \frac{\alpha^{z-1} e^{-q\alpha} }{1-e^{-\alpha}} 
\end{equation}
for the Hurwitz Zeta function, we end up with Eq.~(\ref{isolimit}). This
equation confirms that the complicated expression~(\ref{delani2}) for the
canonical condensate fluctuation in a two-dimensional anisotropic harmonic
trap indeed becomes equal to the expression~(\ref{FLD2}) in the isotropic
limit.

\section{Moments of partitions}
\label{AC}

The saddle-point method followed in Section~\ref{S4} can be employed
to derive asymptotic expressions for the $k$-th moments $\mu_k(n)$ of
unrestricted partitions of integer~$n$, for arbitrary~$k$~\cite{Richmond75}.
Defining the symbol
\begin{equation}
\sum\left[\lambda_0,\lambda_1,\ldots,\lambda_{k-1}\right] \; \equiv \;
\sum\frac{k!}{\ell_1!\,\ell_2!\,\ldots\,\ell_k!}
    \left(\frac{\lambda_0}{1!}\right)^{\ell_1} \ldots
    \left(\frac{\lambda_{k-1}}{k!}\right)^{\ell_k}  \; ,
\end{equation} 
where the sum extends over all partitions of~$k$, i.e.,
$\ell_1 + 2\ell_2 + \ldots + k\ell_k = k$, we find
\begin{equation}
\mu_k(n) \; \sim \; \frac{1}{4\sqrt{3}n}
    \exp\!\left(\pi\sqrt{\frac{2n}{3}}\right)
    \left(\frac{\sqrt{6n}}{\pi}\right)^k
    \sum\left[\ln\!\left(\frac{\sqrt{6n}}{\pi}\right) + \gamma,
    \zeta(2), 2!\,\zeta(3), \ldots , (k-1)!\,\zeta(k) \right] \; .
\label{GMFO}
\end{equation}
For $k = 0$, this expression gives the Hardy--Ramanujan formula~(\ref{HRPN});
for $k = 1$, $2$, and $3$, it adopts the forms
\begin{eqnarray}
\mu_1(n) & \sim & \mu_0(n) \, \frac{\sqrt{6n}}{\pi} \,
    \left[ \ln\!\left(\frac{\sqrt{6n}}{\pi}\right) + \gamma \right]
    \; , \label{MU1N}	\\
\mu_2(n) & \sim & \mu_0(n)	\left(\frac{\sqrt{6n}}{\pi}\right)^2
    \left[ \left(\ln\!\left(\frac{\sqrt{6n}}{\pi}\right) + \gamma\right)^{\!2}
    + \zeta(2) \right]
    \; , \label{MU2N}	\\
\mu_3(n) & \sim & \mu_0(n) \left(\frac{\sqrt{6n}}{\pi}\right)^3
    \left[ \left(\ln\!\left(\frac{\sqrt{6n}}{\pi}\right) + \gamma\right)^{\!3}
    + 3\left(\ln\!\left(\frac{\sqrt{6n}}{\pi}\right) + \gamma\right)\zeta(2)
    + 2\zeta(3) \right]
    \; . \label{MU3N}
\end{eqnarray}
Note that our result~(\ref{GMFO}) differs for $k \ge 2$ from the formula
stated by Richmond~\cite{Richmond75}, and remedies the discrepancies
found by this author when comparing his formula with exact numerical
data. In fact, the above expressions are fairly accurate; some exact
values of $\mu_k(n)$ for $k = 0$ to $3$ are iuxtaposed in Tables~I to~IV
to the respective asymptotic predictions. For completeness, exact values
of the r.m.s.-fluctuation $\sigma(n)$ of the number of parts occuring in
unrestricted partitions of $n$ are listed in Table~V, together with the
approximation furnished by Eq.~(\ref{MFI1}). Comparing the numbers in this
table to those in Table~I, one gets a vivid impression what it means to
isolate microcanonical fluctuations from an exponentially large background.

\begin{figure}
\caption[FIG.~1] {R.m.s.-fluctuation $\sigma(n)$ of the number of integer
    summands into which the integer $n$ can be partitioned. The upper dashed
    line is the leading approximation $\sigma(n) \sim \sqrt{n}$; the lower
    dashed line (coinciding almost with the full line) is the more accurate
    approximation obtained from the square root of Eq.~(\ref{MFI1}).
    The full line indicates the exact values.
    Some numerical data are listed in Table~V of Appendix~\ref{AC}.
    From the viewpoint of statistical mechanics, the upper dashed line gives
    the r.m.s.-fluctuation of the number of ground state particles for an
    ideal Bose gas in a one-dimensional harmonic oscillator trap kept in
    contact with some heat bath, such that the average number~$n$
    of excitation quanta does not exceed the particle number. The other two
    lines correspond to the (approximate and exact) microcanonical condensate
    fluctuation, that is, to the r.m.s.-fluctuation of the number of ground
    state particles when the gas is totally isolated from its surrounding,
    carrying $n$ excitation quanta.} 
\end{figure}

\begin{table}
\begin{tabular}{||c|c|c|c||}  \hline
$n$  &  $\mu_0(n)$ (exact)         &  $\mu_0(n)$ (asymptotic)    &
    rel.\ error	        \\    \hline
50   &  $0.2042260 \cdot 10^6$     &  $0.2175905 \cdot 10^6$     & 0.0654 \\
100  &  $0.1905693 \cdot 10^9$     &  $0.1992809 \cdot 10^9$     & 0.0457 \\ 
200  &  $0.3972999 \cdot 10^{13}$  &  $0.4100251 \cdot 10^{13}$  & 0.0320 \\
300  &  $0.9253083 \cdot 10^{16}$  &  $0.9494095 \cdot 10^{16}$  & 0.0260 \\
500  &  $0.2300165 \cdot 10^{22}$  &  $0.2346387 \cdot 10^{22}$  & 0.0201 \\
1000 &  $0.2406147 \cdot 10^{32}$  &  $0.2440200 \cdot 10^{32}$  & 0.0142 \\
1500 &  $0.1329462 \cdot 10^{40}$  &  $0.1344797 \cdot 10^{40}$  & 0.0115 \\
\hline
\end{tabular}
\caption[Table~1.]{Comparison of exact numbers $\mu_0(n)$ of
    unrestricted partitions of $n$ with the Hardy--Ramanujan
    approximation~(\ref{HRPN}).}
\end{table}

\begin{table}
\begin{tabular}{||c|c|c|c||}  \hline
$n$  &  $\mu_1(n)$ (exact)         &  $\mu_1(n)$ (asymptotic)    &
    rel.\ error         \\    \hline
50   &  $0.2805218 \cdot 10^7$     &  $0.2740428 \cdot 10^7 $    & 0.0231 \\
100  &  $0.4144913 \cdot 10^{10}$  &  $0.4087936 \cdot 10^{10}$  & 0.0137 \\
200  &  $0.1357412 \cdot 10^{15}$  &  $0.1346191 \cdot 10^{15}$  & 0.0083 \\
300  &  $0.4102848 \cdot 10^{18}$  &  $0.4077577 \cdot 10^{18}$  & 0.0062 \\
500  &  $0.1411488 \cdot 10^{24}$  &  $0.1405470 \cdot 10^{24}$  & 0.0043 \\
1000 &  $0.2281551 \cdot 10^{34}$  &  $0.2275624 \cdot 10^{34}$  & 0.0026 \\
1500 &  $0.1621438 \cdot 10^{42}$  &  $0.1618281 \cdot 10^{42}$  & 0.0019 \\
\hline
\end{tabular}
\caption[Table~2.]{Comparison of exact first moments $\mu_1(n)$
    of unrestricted partitions of $n$ with the asymptotic
    formula~(\ref{MU1N}).}
\end{table}
				
\begin{table}
\begin{tabular}{||c|c|c|c||}  \hline
$n$  &  $\mu_2(n)$ (exact)         &  $\mu_2(n)$ (asymptotic)    &
    rel.\ error         \\    \hline
50   &  $0.4461898 \cdot 10^8$     &  $0.4539366 \cdot 10^8$     & 0.0174 \\
100  &  $0.1027721 \cdot 10^{12}$  &  $0.1037857 \cdot 10^{12}$  & 0.0099 \\
200  &  $0.5209742 \cdot 10^{16}$  &  $0.5239850 \cdot 10^{16}$  & 0.0058 \\
300  &  $0.2027390 \cdot 10^{20}$  &  $0.2036083 \cdot 10^{20}$  & 0.0043 \\
500  &  $0.9563321 \cdot 10^{25}$  &  $0.9591871 \cdot 10^{25}$  & 0.0030 \\
1000 &  $0.2361756 \cdot 10^{36}$  &  $0.2366168 \cdot 10^{36}$  & 0.0019 \\
1500 &  $0.2146020 \cdot 10^{44}$  &  $0.2149100 \cdot 10^{44}$  & 0.0014 \\
\hline
\end{tabular} 
\caption[Table~3.]{Comparison of exact second moments $\mu_2(n)$
    of unrestricted partitions of $n$ with the asymptotic
    formula~(\ref{MU2N}).}
\end{table}

\begin{table}
\begin{tabular}{||c|c|c|c||}  \hline
$n$  &  $\mu_3(n)$ (exact)         &  $\mu_3(n)$ (asymptotic)    &
    rel.\ error         \\    \hline
50   &  $0.8145597 \cdot 10^9$     &  $0.9334154 \cdot 10^9$     & 0.1459 \\
100  &  $0.2898292 \cdot 10^{13}$  &  $0.3173679 \cdot 10^{13}$  & 0.0950 \\
200  &  $0.2249985 \cdot 10^{18}$  &  $0.2390975 \cdot 10^{18}$  & 0.0627 \\
300  &  $0.1120055 \cdot 10^{22}$  &  $0.1175340 \cdot 10^{22}$  & 0.0494 \\
500  &  $0.7186145 \cdot 10^{27}$  &  $0.7449881 \cdot 10^{27}$  & 0.0367 \\
1000 &  $0.2683336 \cdot 10^{38}$  &  $0.2749645 \cdot 10^{38}$  & 0.0247 \\
1500 &  $0.3099702 \cdot 10^{46}$  &  $0.3160662 \cdot 10^{46}$  & 0.0197 \\
\hline
\end{tabular}
\caption[Table~4.]{Comparison of exact third moments $\mu_3(n)$
    of unrestricted partitions of $n$ with the asymptotic
    formula~(\ref{MU3N}).}
\end{table}
	
\begin{table}
\label{T5}
\begin{tabular}{||c|c|c|c||}  \hline
$n$  &  $\sigma(n)$ (exact)  &  $\sigma(n)$ (asymptotic)  & rel.\ error \\
\hline
50   &  5.46  &  5.65  &  0.0349    \\
100  &  8.14  &  8.29  &  0.0190    \\
200  & 12.00  & 12.12  &  0.0104    \\
300  & 15.00  & 15.11  &  0.0073    \\
500  & 19.80  & 19.89  &  0.0047    \\
1000 & 28.71  & 28.79  &  0.0026    \\
1500 & 35.60  & 35.66  &  0.0018    \\
\hline
\end{tabular}
\caption[Table~5.]{Comparison of exact r.m.s.-fluctuations $\sigma(n)$
    of the number of parts in unrestricted partitions of $n$ with the
    predictions obtained by taking the square root of the asymptotic
    Eq.~(\ref{MFI1}).}
\end{table}
	    	
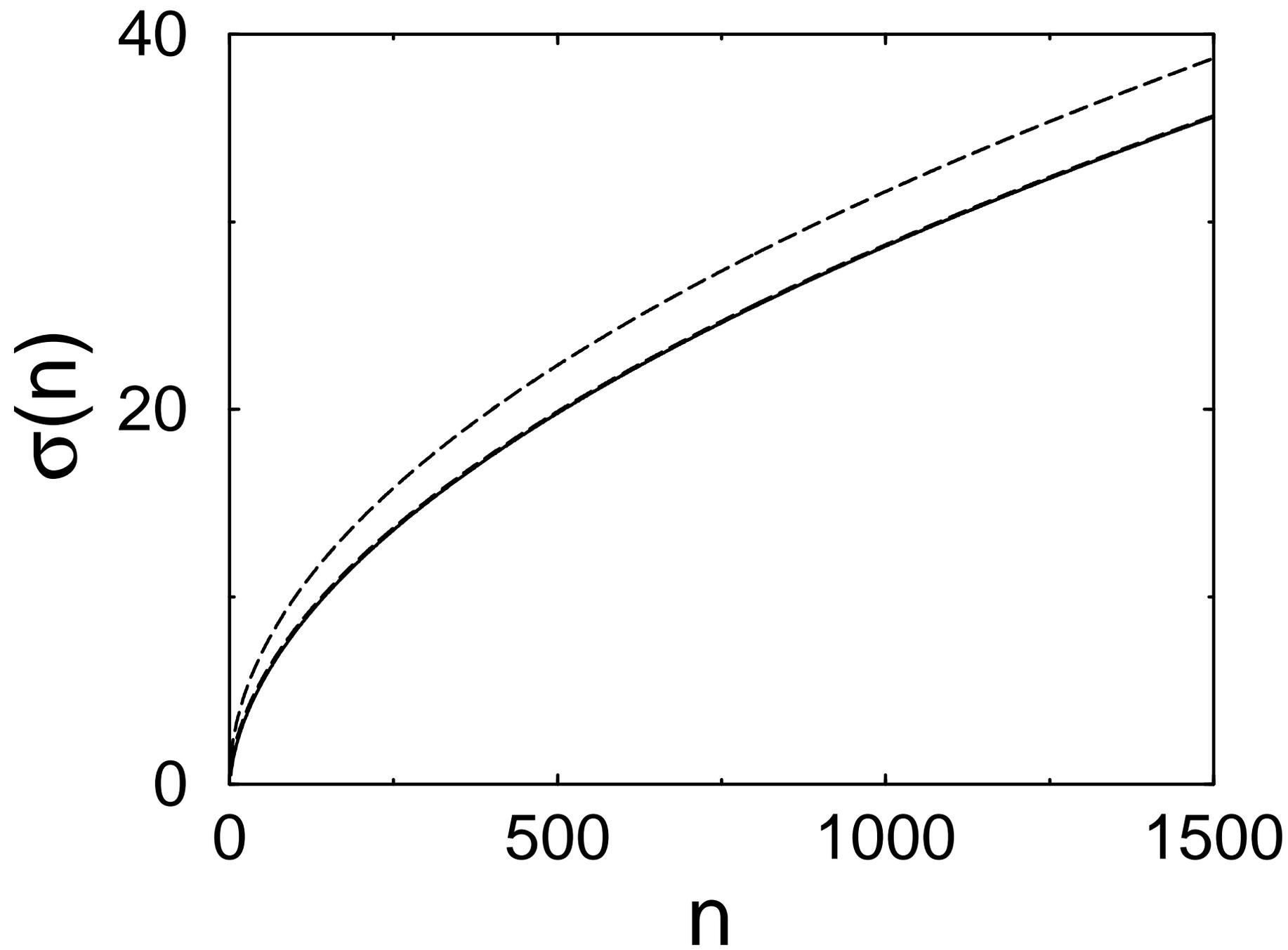
\end{document}